\pdfoutput=1
\newif\ifHighlight
\newif\ifFullVersion

\FullVersiontrue
\Highlightfalse

\documentclass[acmsmall,screen]{acmart}

\AtBeginDocument{%
  \providecommand\BibTeX{{%
    \normalfont B\kern-0.5em{\scshape i\kern-0.25em b}\kern-0.8em\TeX}}}

\setcopyright{rightsretained}
\acmPrice{}
\ifFullVersion
\acmDOI{nn.nnnn/nnnnnnn}
\else
\acmDOI{10.1145/3485482}
\fi
\acmYear{2021}
\copyrightyear{2021}
\acmSubmissionID{oopsla21main-p32-p}
\acmJournal{PACMPL}
\acmVolume{5}
\acmNumber{OOPSLA}
\acmArticle{105}
\acmMonth{10}

\setcitestyle{round,comma}

\usepackage{alltt}
\usepackage{stmaryrd}
\usepackage{multirow}
\usepackage{listings}
\usepackage{fancyvrb}
\newcommand{\SysName}{{\sc Synbit}}

\input{kztk_macro}
\newcommand{\Hole}{\Box}

\newcommand{\EHoleS}[1]{\Hole^\mathrm{e}(#1)}
\newcommand{\RHoleS}[2]{\Hole^\mathrm{r}(#1,#2)}
\newcommand{\GHole}{\Hole}
\newcommand{\SHole}[1]{\Hole(#1)}

\newcommand{\EHole}[3]{\Hole^\mathrm{e}(#1,#2;#3)}
\newcommand{\RHole}[4]{\Hole^\mathrm{r}(#1,#2;#3,#4)}

\newcommand{\EvalRelFT}[4]{#1 \vdash_\mathrm{GT} #2 \evalF_{#4} #3}
\newcommand{\EvalRelBT}[5]{#1 \vdash_\mathrm{PT} #2 \evalB_{#4} #3 \dashv #5}

\newcommand{\TrNil}{\epsilon}
\newcommand{\TrBranch}[3]{\mathrm{Br}(#1,#2,#3)}

\newcommand{\GenTmp}[7]{#1; #2; #3 \triangleright #4\vdash #5 : #6 \leadsto #7}

\makeatletter
\AtBeginDocument{\let\c@lstlisting\c@figure}
\makeatother

{}
\definecolor{darkpastelgreen}{rgb}{0.01, 0.75, 0.24}
\definecolor{seagreen}{rgb}{0.18, 0.55, 0.34}
\definecolor{richlavender}{rgb}{0.67, 0.38, 0.8}
\definecolor{darkcyan}{rgb}{0.0, 0.55, 0.55}
\definecolor{jade}{rgb}{0.0, 0.66, 0.42}
\definecolor{caribbeangreen}{rgb}{0.0, 0.8, 0.6}

\ifHighlight
\newcommand{\revised}[1]{\textcolor{darkpastelgreen}{#1}}

\newenvironment{revisedblock}{%
  \color{darkpastelgreen}%
}{%
  \ignorespacesafterend%
}
\else
\newcommand{\revised}[1]{#1}
\newenvironment{revisedblock}{}{}
\fi

\makeatletter
\@ifundefined{mathindent}%
  {\newdimen\mathindent\mathindent\leftmargini}%
  {}%
\makeatother
\newlength{\blanklineskip}
\makeatletter
\newenvironment{codemath}{%
    \start@align\tw@\st@rredtrue\m@ne%
    \hskip\mathindent%
}{%
    \endalign
}
\makeatother

\begin{document}

\title{\SysName: Synthesizing Bidirectional Programs \revised{using} Unidirectional Sketches}

\author{Masaomi Yamaguchi}
\email{masaomi.yamaguchi.t4@dc.tohoku.ac.jp}
\orcid{0000-0002-7347-6021}
\affiliation{%
  \department{Graduate School of Information Sciences}
  \institution{Tohoku University}
  \city{Sendai}
  \state{Miyagi}
  \country{Japan}
}
\authornote{Currently at Fujitsu.}

\author{Kazutaka Matsuda}
\orcid{0000-0002-9747-4899}
\email{kztk@ecei.tohoku.ac.jp}
\affiliation{%
  \department{Graduate School of Information Sciences}
  \institution{Tohoku University}
  \city{Sendai}
  \state{Miyagi}
  \country{Japan}
}

\author{Cristina David}
\orcid{0000-0002-9106-934X}
\email{cristina.david@bristol.ac.uk}
\affiliation{%
  \institution{University of Bristol}
  \streetaddress{BS8 1QU}
  \city{Bristol}
  \state{Avon}
  \country{UK}
}

\author{Meng Wang}
\orcid{https://orcid.org/0000-0001-7780-630X}
\email{meng.wang@bristol.ac.uk}
\affiliation{%
  \institution{University of Bristol}
  \streetaddress{BS8 1QU}
  \city{Bristol}
  \state{Avon}
  \country{UK}
}



\begin{abstract}
  We propose a technique for synthesizing bidirectional programs from
  the corresponding unidirectional code plus a few input/output
  examples.  The core ideas are: \revised{(1) \emph{constructing a sketch} using the given unidirectional program as a specification, and (2) \emph{filling the
  sketch} in a modular fashion by exploiting the properties of bidirectional programs.} 
  These ideas are enabled by our choice of
  programming language, HOBiT, which is specifically designed to
  \revised{maintain the unidirectional program structure in bidirectional programming, 
  and keep the parts that control bidirectional behavior modular.} 
  To evaluate our
  approach, we implemented it in a tool called {\SysName} and used it
  to generate bidirectional programs for intricate microbenchmarks, as
  well as for a few larger, more realistic problems.  We also compared
    {\SysName} to a state-of-the-art unidirectional synthesis tool on the task of
  synthesizing backward computations.

\end{abstract}



\maketitle

\section{Introduction}\label{sec:intro}

Transforming data from one format to another is a common task of
programming: compilers transform program text into syntax trees,
manipulate the trees and then generate low-level code; database
queries transform base relations into views; model-driving software engineering transforms one model into another. Very often,
such transformations will benefit from being bidirectional, allowing
changes to the targets to be mapped back to the sources too (for example the
view-update problem in databases~\cite{BaSp81,Hegner90}, bidirectional model transformation~\cite{Stevens2008}, and so on).

As a
response to this need, programming-language researchers started to design
specialized programming languages for writing bidirectional transformations. In particular as
pioneered by Pierce's group at Pennsylvania, a \emph{bidirectional transformation} (BX), also known as a \emph{lens}~\cite{FGMPS07}, is modeled as a pair of functions
between source and view data objects, one in each direction. The forward function $\GET::S \to V$ maps
a source onto a view, and the corresponding backward function
$\PUT:: S \times V \to S$ reflects any changes in the view back to the source.
Note that \revised{$\GET$ is not necessarily injective. Accordingly $\PUT$,
 in addition to the updated view,
also takes the original source as an argument. This makes it possible to recover some of the source data
that is not present in the view.} Of course, not all pairing of $\GET$\slash$\PUT$ forms are valid BX; they must be related by specific properties known as \emph{round-tripping}. 
\begin{alignat}{2}
       & \GET \A s = v      &  & \quad\text{implies}\quad \PUT \A (s, v) = s \tag{\textbf{Acceptability}}\label{law:acceptability} \\
       & \PUT \A (s,v) = s' &  & \quad\text{implies}\quad \GET \A s' = v     \tag{\textbf{Consistency}}\label{law:consistency}
\end{alignat}
for all $s, s' \in S$ and $v \in V$. 
Here, \ref{law:acceptability} states that no changes to the source happen if there is no change to the view,
and \ref{law:consistency} states that all changes to the view must be captured in the updated source.

A BX language allows the transformations in both directions to be programmed together and is expected to guarantee round-tripping by construction. 

This is a challenging problem for language design, and consequently compromises had to be made (in particular to usability) in favor of guaranteeing round-tripping. In the original lens design~\cite{FGMPS07}, lenses can only be composed by stylized lens combinators, which is inconvenient to program with. A lot of research has gone into this area since, for example~\citet{MatsudaHNHT07, DBLP:conf/popl/BohannonFPPS08, bff,PaHF14,Matsuda&Wang:2018:HOBiT}, and the state of the art has progressed a long way since. This includes a language HOBiT~\cite{Matsuda&Wang:2018:HOBiT}, which follows a line of research~\cite{bff, DBLP:conf/icfp/MatsudaW15, DBLP:journals/scp/MatsudaW15, MatsudaHNHT07} that aims to produce BX code that is close in structure to how one will program the $\GET$ function alone in a conventional unidirectional language.
Despite the progresses in language design, BX programming is still considerably more difficult than conventional programming, especially when sophisticated backward behaviors are required. This complexity is largely inherent as one is asked to do more in less: defining behaviors in both directions in a single definition. Even in a language like HOBiT, where programmers are allowed (and indeed encouraged) to approach BX programming from the convenience of conventional unidirectional programming, there are still (necessary) additional code components that need to be added to the basic program structure to specify non-trivial backward behaviors.



\paragraph{Unidirectional Program as Sketch}
In this paper we introduce \SysName, a program synthesis system that makes BX programming more approachable to mainstream programmers.
In particular, we propose using unidirectional code (i.e., a definition of $\GET$ in a Haskell-like language)
as a sketch of the bidirectional program (which embodies both $\GET$ and $\PUT$).
\revised{Consequently, programmers familiar with unidirectional programming can obtain bidirectional programs from unidirectional ones and input/output examples.}
In the neighboring field of software verification,
expressing specifications (in our case sketches) as normal code has the effect of boosting the adoption of formal tools
in industry~\cite{DBLP:conf/icse/ChongCKKMSTTT20}, something that bidirectional programming research as a whole may benefit from.



It is not hard to see that this program sketch idea fits well with the
language HOBiT. Unlike most BX languages, HOBiT is designed to keep
bidirectional code as similar in structure as possible to how one may program the unidirectional
$\GET$. Consequently, it is able to benefit from such a sketch and allow
the synthesis process to mostly focus on
parts of the code that specially handle intricate bidirectional behaviors.
This is an attractive solution. On one hand, the specifications are intuitive: users simply write normal unidirectional programs (together with a few input/output examples).
\revised{On the other hand,
the specifications as sketches are useful in the synthesis process because they reduce
the search space.} Moreover, this design supports gradual ``bidirectionalization''
done by incrementally converting existing unidirectional programs into bidirectional ones.
As it will be shown in a comprehensive evaluation in Section~\ref{sec:experiments},
our system is highly effective and able to produce high-quality bidirectional programs in
a wide range of scenarios.

\paragraph{Off-the-shelf synthesis is a non-solution}
Before diving into the details of our proposed solution, we would like to
take a step back and answer a question that may already be in some readers' minds:
will program synthesis completely replace the need for bidirectional languages? That is, how
about using generic synthesizers to derive a $\PUT$ from an existing $\GET$ in a standard
unidirectional language? After all, there already exist bidirectionalization techniques~\cite{bff, MatsudaHNHT07} that
are able to derive a $\PUT$ from a $\GET$ though in restricted situations.

When applied naively, this approach does not work.  As an experiment, we tried
using the state-of-the-art program synthesizer \textsc{Smyth}~\cite{DBLP:journals/pacmpl/LubinCOC20} to generate the
$\PUT$ from concrete examples and appropriate sketches.  To simplify the problem, we ignored
the round-tripping property between $\GET$ and $\PUT$, and tried to
generate any $\PUT$ (even one that violates the laws). However, even
in this simplified scenario, the synthesizer failed to find a $\PUT$
for simple examples (see Section~\ref{sec:experiment-comparative} for more details).

This is not surprising because, while powerful, program synthesis is very
hard due to the vast search space. The most common ways in which existing
synthesis techniques circumvent this are by picking a reduced domain specific language
to generate programs in \cite{DBLP:conf/popl/Gulwani11} and by seeding the program search with a sketch
representing the program structure \cite{DBLP:conf/aplas/Solar-Lezama09}.
In this paper, we are interested in synthesizing general purpose programs
and therefore we do not adopt the first strategy.
%

\paragraph{Contributions:}
\begin{itemize}
\item \revised{We present an application of program synthesis to the area of bidirectional programming.}
            In particular, we provide an
            automated technique for generating bidirectional
            transformations in the language HOBiT (Section~\ref{sec:synthesis}). The inputs to our procedure
            are the corresponding unidirectional code and a few concrete
            examples describing the backward transformation (Section~\ref{sec:synthesis-input}).


      \item We exploit bidirectional programming properties, domain-specific
            knowledge of HOBiT and type information to efficiently prune the
            search space.  In particular, we generate specialized program
            sketches from the unidirectional code (Section~\ref{sec:template-generation}), which are then filled
            in a modular manner by separating the solving of dependent synthesis tasks
            (Sections~\ref{sec:filling-holes} and
            \ref{sec:synthesis-search-and-filtering}).

      \item We present a classification of bidirectional programming
            benchmarks based on the amount of information from the source that
            is being lost through the forward transformation
            (Section~\ref{sec:experiment-categorize}). We believe that such a
            classification is valuable for evaluating the capabilities of our
            bidirectional synthesis technique.



      \item We implemented our bidirectional synthesis technique in a tool
            called {\SysName}, and used it to generate bidirectional
            programs for the set of benchmarks discussed above (Section~\ref{sec:experiments}).
            The prototype implementation of {\SysName} is available in the artifact \footnote{\url{https://doi.org/10.5281/zenodo.5494504}} or the repository\footnote{\url{https://github.com/masaomi-yamaguchi/synbit}}.
\end{itemize}



\section{Background: the HOB{\lowercase{i}}T language}
\label{sec:prelim}

HOBiT~\cite{Matsuda&Wang:2018:HOBiT} is a state-of-the-art higher-order bidirectional programming language.
\revised{A distinct feature of HOBiT is its support of a programming style that is
close to the conventional unidirectional programming.} The design of the language largely separates the core structure
of programs (which can be shared with the unidirectional definition of $\GET$) from the specification of backward behaviors that
are specific to bidirectional programming. In this section, we will introduce the core features of HOBiT with a focus on demonstrating its
suitability as a target of sketch-based program synthesis. Curious readers who are interested in the full expressiveness power of HOBiT and
the formal systems are encouraged to read the original paper~\cite{Matsuda&Wang:2018:HOBiT}.


\subsection{A Simple Example}
Before getting into HOBiT programs, we start with a familiar definition in Haskell below.
\begin{codemath}\bb
  \var{append} :: {[a]} \to {[a]} \to {[a]} \\
  \var{append} \A \var{xs} \A \var{ys} = \CASE~\var{xs}~\OF~
  \bbt
  [\,]  & \to \var{ys} \\
  a : x & \to a : \var{append} \A x \A \var{ys}
  \ee
  \ee&&\end{codemath}
In the definition, we use explicit case branching (instead of syntax sugar in Haskell) to highlight the
structure of the code. 

Now, for a forward function ($\GET$) defined as $\var{append}$, let us investigate what will be suitable behaviors of its $\PUT$.
We denote the $\PUT$ by a HOBiT function $\var{appendB} :: \BX{[a]} \to \BX{[a]} \to \BX{[a]}$. The $\BX{}$-annotated types (highlighted in \textcolor{bxcolor}{blue}) are
\emph{bidirectional types} in HOBiT, representing data that are subject to bidirectional computation. \revised{$\BX{}$-typed values are manipulated only by operations that satisfy the round-tripping laws, which is enough to ensure the round-tripping property of a whole program~\cite{Matsuda&Wang:2018:HOBiT}.} As we will see in the sequel, bidirectional types can be mixed with normal unidirectional types to support flexible programming and greater expressiveness.

Bidirectional functions of type $\BX{\sigma} \to \BX{\tau}$ can be executed as bidirectional transformations between $\sigma$ and $\tau$
in HOBiT's interactive environment (or, read-eval-print loop) via \texttt{:get} and \texttt{:put}.
For example, one can run $\var{appendB}$ forwards
\begin{codemath}\bb
  \texttt{>}~\texttt{:get} \A (\var{uncurryB} \A \var{appendB}) \A ([1,2], [3,4])  \\
  \ \!\! [1,2,3,4]
  \ee&&\end{codemath}
and backwards.
\begin{codemath}\bb
  \texttt{>}~\texttt{:put} \A (\var{uncurryB} \A \var{appendB}) \A ([1,2], [3,4]) \A [5,6,7,8] \\
  ([5,6], [7,8])
  \ee&&\end{codemath}
Note that we have \emph{uncurried} $\var{appendB}$ before execution by $\var{uncurryB} :: (\BX{a} \to \BX{b} \to \BX{c}) \to \BX{(a,b)} \to \BX{c}$ so that it fits the pattern of
$\BX{\sigma} \to \BX{\tau}$ for bidirectional execution. Specifically $(\var{uncurryB} \A \var{appendB})$ has type $\BX{([a],[a])} \to \BX{[a]}$, and its $\PUT$ has type $([a],[a])\to [a] \to ([a],[a])$.

Now we are ready to explore bidirectional behaviors.

\subsubsection{Simple Backward Behavior} The simplest behavior of $\PUT$, as adopted in~\citet{bff}, is to only
allow \emph{in-place} update of views. In the case of $\var{appendB}$, it means that the changes to the length of
the view list will result in an error.
\begin{codemath}\bb
  \texttt{>}~\texttt{:put} \A (\var{uncurryB} \A \var{appendB}) \A ([1,2], [3,4]) \A [5,6,7,8] \\
  ([5,6], [7,8])\\
  \texttt{>}~\texttt{:put} \A (\var{uncurryB} \A \var{appendB}) \A ([1,2], [3,4]) \A [1,2,3] \\
  \texttt{Error: ...}
  \ee&&\end{codemath}
To achieve this behavior, a definition in HOBiT reads the following.
\begin{codemath}\bb
  \var{appendB} :: \BX{[a]} \to \BX{[a]} \to \BX{[a]} \\
  \var{appendB} \A \var{xs} \A \var{ys} = \BCASE~\var{xs}~\BOF~
  \bbt
  [\,]  & \to \var{ys} && \\
  a : x & \to a \bxop{:} \var{appendB} \A x \A \var{ys} & & 
  \ee
  \ee&&\end{codemath}
As one can see, this definition is almost identical to that of $\var{append}$ with only the language constructs
such as case and data constructors being replaced by their bidirectional counterparts (underlined and highlighted in \textcolor{bxcolor}{blue}) that handle values of bidirectional types.

\revised{This simplicity comes from the design of HOBiT, as well as the modesty of the scenario.} Given that the function is parametric in the list elements, in-place updates mean that the backward execution may simply trace back
exactly the same control flow of the original forward execution. This can be achieved by recursing according to the original source (the first argument of $\PUT$) and only using the updated view (the second argument of $\PUT$) as a supplier of element values.
Therefore, no additional specification is required in the code.

\subsubsection{Branch Switching}
HOBiT is not limited to such simple behaviors. Its bidirectional language constructs seen above set us up for more sophisticated cases.
Let's say that we now want to handle structural updates in the view, allowing the list length to vary.
\begin{codemath}\bb
  \texttt{>}~\texttt{:get} \A (\var{uncurryB} \A \var{appendB}) \A ([1,2], [3,4])  \\
  \ \!\! [1,2,3,4]\\
  \texttt{>}~\texttt{:put} \A (\var{uncurryB} \A \var{appendB}) \A ([1,2], [3,4]) \A [5,6,7,8] \\
  ([5,6], [7,8]) \\
  \texttt{>}~\texttt{:put} \A (\var{uncurryB} \A \var{appendB}) \A ([1,2], [3,4]) \A [5,6,7,8,9] \\
  ([5,6], [7,8,9]) \\
  \texttt{>}~\texttt{:put} \A (\var{uncurryB} \A \var{appendB}) \A ([1,2], [3,4]) \A [5] \\
  ([5], [\,])
  \ee&&\end{codemath}
When the length of the view list changes, we try to change the second list of the source to accommodate that. If the length becomes shorter than that of the first source list, the second source list will be empty and the first source list will also change accordingly.

As one can see, this behavior can no longer be achieved by simply tracing back the original control flow of the forward execution.
The backward execution will have to recurse a different number of times from the original, and how this is done will need to
be additionally specified in the code. Here enters a definition in HOBiT that does exactly this.
\begin{codemath}\bb
  \var{appendB} :: \BX{[a]} \to \BX{[a]} \to \BX{[a]} \\
  \var{appendB} \A \var{xs} \A \var{ys} = \BCASE~\var{xs}~\BOF~
  \bbt
  [\,]  & \to \var{ys} & \BWITH~\var{const} \A \con{True} & \BBY~\lambda \dontcare.\lambda \dontcare.\, [\,] \\
  a : x & \to a \bxop{:} \var{appendB} \A x \A \var{ys} & \BWITH~\var{not} \circ \var{null} & \BBY~\lambda s.\lambda \dontcare.\, s 
  \ee
  \ee&&\end{codemath}
The code is longer than the last version, as expected, but the program structure
remains the same: the additional specification for more sophisticated backward behavior is modularly grouped at the end of
each case branch. 
\revised{Recall that we plan to use the
unidirectional code as sketches to synthesize bidirectional code; this resemblance to the unidirectional code means that the synthesizing effort may now
concentrate on the part specifying bidirectional behaviors, increasing its effectiveness.}

In the above code, we used two distinctive HOBiT features known as \emph{exit conditions} (marked by the $\BWITH$ keyword) and \emph{reconciliation functions} (marked by the $\BBY$ keyword). Both are for the purpose of controlling the backward behavior, especially when it no longer follows the original control flow (a behavior we call \emph{branch switching}).

\paragraph {\bf Exit conditions.} An exit condition is an over-approximation of the forward-execution result of the branch, which
always evaluates to True if the branch is taken (dynamically checked in HOBiT). \revised{Hence, an exit condition in a $\BCASE$ expression has type $\tau \to \con{Bool}$ if the whole $\BCASE$ expression has type $\BX{\tau}$.} The exit conditions are then used
as branching conditions in the backward execution. For example, in the above case of \var{appendB}, an empty list as view will
choose the first branch, as the view does not match the condition $\var{not} \circ \var{null}$ of the second branch.
Exit conditions often overlap; when multiple branches match, the original branch used in the forward execution is preferred. If
impossible (as the exit condition of that branch does not hold), the topmost branch will be taken. Like the case of the in-place update we saw previously, if the view-list length is not changed, then in the backward execution of \var{appendB}, the exit conditions of the original branches (now used as branching conditions) are always satisfied, and therefore the original branches are always taken.

The situation becomes more interesting when the view update \emph{does} change the length of the list, for example by making it shorter.
In this case, the view list will be exhausted before the original number of recursions are completed. As a result, the backward
execution will now see $[\,]$ as its view input and a non-empty list as its source input. This means that the original branch at this
point is the second branch, but the exit condition of that does not hold, which forces the first branch to be taken---a \emph{branch switch}.

\paragraph {\bf Reconciliation functions.} We have seen that exit conditions may force branches to switch, which is crucial for handling interesting changes to the view. However, it only solves half of the problem; naive branch switching typically results in run-time failure. The reason is simple: when branch switching happens, the two arguments of $\PUT$ are in an inconsistent state for the branch;
\eg, for $\var{append}$, having an non-empty source list (and an empty view list) is inconsistent for the branch $[] \to \var{ys}$.
Reconciliation functions are used to fix this inconsistency. Basically, they are functions that take the inconsistent sources and views and produce new sources that are consistent with the branch taken. For example, in the definition above, the first branch will have $[\,]$ as the new source, because a switch to this branch means an empty view 
and the branch expects the source to be the empty list for further $\PUT$ execution of the branch body.
\revised{In general, a reconciliation function in a $\BCASE$ expression is a function of type $\sigma \to \tau \to \sigma$, provided that the whole $\BCASE$ expression has type $\BX{\tau}$, with its scrutinee of type $\BX{\sigma}$.}

An interesting observation of this particular example of $\var{append}$ is that the reconciliation function of the cons branch (\ie, $\lambda s. \lambda \dontcare.s$ above) is actually never used. Recall that branch switching only happens when the backward execution tries to follow the original branch but the exit condition of the branch is not satisfied by the updates to the view. This will never happen in the nil branch above with the exit condition $\var{const} \A \con{True}$, which is always satisfied. In other words, regardless of the view update there will not be branch switching to the cons branch and therefore its reconciliation function is never executed. This behavior matches the behavior of $\var{append}$ which recurses on the first source list: when the view list is updated to be shorter than the first source list, the recursion will need to be cut short (thus branch switching to the nil branch); but when the view list is updated to be longer, the additional elements will simply be added to the second source list, which does not affect the recursion (and thus no need of branch switching).

In summary, with reconciliation functions, the backward execution may recover from inconsistent states and resume with a new source.
This is key to successful branch switching and the handling of structural updates to the view.

\begin{revisedblock}
  \paragraph{\bf Round-tripping}
It is also worth noting that branch switching in HOBiT does not threaten the round-tripping properties.
Intuitively, the key principle of round-tripping is that a branch taken in a forward/backward execution should also be
taken in a subsequent backward/forward execution~\cite{FGMPS07,Janus,YoAG11,KoZH16,DBLP:conf/ac/HuK16,Matsuda&Wang:2018:HOBiT}.
When a brach switches in the backward execution, the new branch will produce a source value that matches the pattern of the new branch,
ensuring that a subsequent forward execution will take the same branch. Since the exit conditions are checked as valid post conditions, this correspondence of forward/backward branchings is established, and consequently it guarantees round-tripping.
An inappropriate reconciliation function will
make the backward execution fail but not break round-tripping.
More details can be found in the original paper~\cite{Matsuda&Wang:2018:HOBiT}.
In this paper, we not only rely on the fact that HOBiT programs always satisfy round-tripping,
but we also leverage the principle for effective synthesis (Section~\ref{sec:synthesis-prefiltering}).




\end{revisedblock}

One can also observe that the exit conditions and reconciliation functions in $\var{appendB}$ are quite simple themselves. However, their interaction with the
rest of the code is intricate. Programmers who write them are therefore required to have a good understanding of how backward execution works and how it can be influenced, which may not come naturally. This combination of simplicity in form and complication in behavior
makes it a fertile ground for program synthesis, which we set out to explore in this paper.

\subsubsection{Mixing Bidirectional and Unidirectional Programming}
We end this section with another example of variants of $\var{append}$'s backward behavior and its implementation in HOBiT.
\revised{The example also demonstrates a feature of HOBiT that supports a mixture of unidirectional and bidirectional programming
for greater expressiveness.} Let us consider the following definition.
\begin{codemath}\bb
  \var{appendBc} :: \BX{[a]} \to [a] \to \BX{[a]} \\
  \var{appendBc} \A \var{xs} \A \var{ys} = \BCASE~\var{xs}~\BOF~\\
  \qquad \bbt
  [\,]  & \to \LIFTU{\var{ys}} & \BWITH~\lambda v.\, \var{length} \A v \doubleequals \var{length} \A \var{ys} & \BBY~\lambda \dontcare.\lambda \dontcare.\, [\,] \\
  a : x & \to a \bxop{:} \var{appendBc} \A x \A \var{ys} & \BWITH~\lambda v.\, \var{length} \A v \not\doubleequals \var{length} \A \var{ys} & \BBY~\lambda \dontcare. \lambda (v:\dontcare).\, [v]
  \ee
  \ee&&\end{codemath}
Noticeably, the type of the function is a mixture of bidirectional and unidirectional types, with the second argument as a normal list.
Recall that bidirectional types represent data that are updatable; this type means that the second list is fixed with respect to
backward execution. We will look at a few sample runs before going into the details of the definition. Note that since the second argument is constant in backward execution, there is no longer the need to uncurry the function; one can simply partially apply it as shown below.
\begin{codemath}\bb
  \texttt{>}~\texttt{:get} \A (\lambda \var{xs}.\, \var{appendBc} \A \var{xs} \A \texttt{";"}) \A \texttt{"apple"}  \\\relax
  \texttt{"apple;"} \\
  \texttt{>}~\texttt{:put} \A (\lambda \var{xs}.\, \var{appendBc} \A \var{xs} \A \texttt{";"}) \A \texttt{"apple"} \A \texttt{"pineapple;"} \\\relax
  \texttt{"pineapple"} \\
  \texttt{>}~\texttt{:put} \A (\lambda \var{xs}.\, \var{appendBc} \A \var{xs} \A \texttt{";"}) \A \texttt{"apple"} \A \texttt{"plum;"} \\\relax
  \texttt{"plum"} \\
  \ee&&\end{codemath}
In this case, the second list is $\texttt{";"}$ and changes in the view can only affect the first list.
Any attempt to change the last part of the view will (rightly) fail.
\begin{codemath}\bb
  \texttt{>}~\texttt{:put} \A (\lambda \var{xs}.\, \var{appendBc} \A \var{xs} \A \texttt{";"}) \A \texttt{"apple"} \A \texttt{"apple."} \\\relax
  \texttt{Error: ...}
  \ee&&\end{codemath}
Now let us go back to the definition. The fact that the second argument $\var{ys}$ is now of a normal (non-bidirectional) type means that it can be used
in the exit conditions and reconciliation functions (which only involve unidirectional terms). During backward execution, the exit conditions dictate that the recursion will terminate (the first branch taken) when the view list is the same length as the original $\var{ys}$. In addition, since $\var{ys}$ has a normal type, it will need to be \emph{lifted} (as a constant) to the bidirectional world by \LIFTU{} so that the $\BCASE$ expression
becomes well typed. 
We again refer interested readers to~\citet{Matsuda&Wang:2018:HOBiT} for lifting in more general forms.

The mixture of unidirectional and bidirectional programming is a challenge to program synthesis as the search space has become much larger. Still, the fundamental has not changed: a definition of $\GET$ remains a good sketch for HOBiT programs.

\section{Synthesis of HOBiT Programs using Unidirectional Programs as Sketches}\label{sec:synthesis}

In this section, we describe our technique for synthesizing bidirectional programs in HOBiT. Throughout the section, we will use the familiar case of
\var{append} as the running example.

\subsection{Overview}\label{sec:overview}
Before presenting the technical details, we start with an informal overview of the synthesis process.
\SysName~takes in a unidirectional program (written in a subset of Haskell) and a small number of
input/output examples of the required backward behavior, and produces a HOBiT program that behaves like the input unidirectional program in the forward direction and is guaranteed to satisfy the round-tripping laws and conform to the given examples in the backward direction. More details on the guarantees of our system are given later in Section~\ref{sec:soundness-incompleteness}.

As an example, in the case of \var{append}, we provide the following specification to \SysName.
\bgroup
\setlength{\jot}{0pt}
\begin{align*}
   & \var{append} :: [\con{Int}] \to [\con{Int}] \to [\con{Int}]                                                                                                       \\
   & \var{append}                 = \lambda \var{xs}.\, \lambda \var{ys}.\, \CASE~\var{xs}~\OF~\{ [\,] \to \var{ys} ;\; (a:x) \to a : \var{append} \A x \A \var{ys} \} \\[\blanklineskip]%
   & \texttt{:put} \A (\var{uncurryB} \A \var{appendB})  \A ([1,2,3], [4,5]) \A  [6,2]  = ([6,2], [\,])
\end{align*}
\egroup%
The definition of \var{append} above is completely standard. The user-provided input/output example specifies that the view list may be updated
to a smaller length. As we have seen in Section~\ref{sec:prelim}, \var{append} needs to be
uncurried before bidirectional execution, which is also reflected in the input/output example above where the source is a pair
of lists. One interesting observation is that this bidirectional execution provides a call context of the function to be
synthesized, which speeds up the synthesis
process by narrowing down the choices of \var{appendB}'s type.

For the given specification, \SysName~produces the following result.
\bgroup
\setlength{\jot}{0pt}
\begin{codemath}\bb
  \var{appendB} :: \BX{[\con{Int}]} \to \BX{[\con{Int}]} \to \BX{[\con{Int}]}\\
  \var{appendB} = \lambda \var{xs}.\, \lambda \var{ys}.\, \BCASE~\var{xs}~\BOF~\{
  \begin{tarray}[t]{lll}
    [\,] \to \var{ys} \\
    \qquad\begin{tarray}[t]{ll}
      \BWITH &\lambda v.\,\CASE~v~\OF~\{ x \to \con{True};\; \dontcare \to \con{False} \}\\
      \BBY   &\lambda s. \lambda v.\,\CASE~v~\OF~\{ x \to  [\,]
      \};
    \end{tarray}
    \\
    (a:x) \to a : \var{append} \A x \A \var{ys}
    \\
    \qquad \begin{tarray}[t]{lll}
      \BWITH &\lambda v.\,\CASE~v~\OF~\{ z : \var{zs}\to \con{True};\; \dontcare \to \con{False} \}\\
      \BBY   &\lambda s. \lambda v.\,\CASE~v~\OF~\{ z : \var{zs} \to 
      s
      \}\}
    \end{tarray}
  \end{tarray}
  \ee&&\end{codemath}
\egroup

%
%
As one can see, this program is equivalent to the hand-written definition in Section~\ref{sec:prelim};
the only difference is that the synthesized version does not use library functions such as \var{const} and \var{null}.\footnote{Obvious cosmetic simplification could be made to part of the code for readability. But that is an orthogonal concern.}

Roughly speaking, the synthesis process that produces the above result involves two major components: the generation of
a suitable sketch with holes and the filling of the holes. We will look at the main steps below.

%
\paragraph{{\bf Generation of sketches}}
The sketch is expected to be largely similar in structure to the unidirectional definition (thanks to the design of HOBiT),
but there are a few details to be ironed out. First of all, one needs to decide the type of the target function.
Recall that HOBiT is a powerful language that supports the mixing of unidirectional and bidirectional programming.
Thus, for a type such as \var{append}'s, there are several possibilities such as $\BX{[\con{Int}]} \to \BX{[\con{Int}]} \to \BX{[\con{Int}]}$, $\BX{[\con{Int}]} \to [\con{Int}] \to \BX{[\con{Int}]}$, $[\BX{\con{Int}}] \to \BX{[\con{Int}]} \to \BX{[\con{Int}]}$, and so on. It is therefore crucial to narrow down the choices to control the search space. The call context in the input/output example(s) in the specification
is useful for this step, as it can effectively restrict its type. 
We will discuss more
details on this in Section~\ref{sec:synthesis-template-generation}. For now, it is sufficient to know that for the specification given in this example, the only
viable type is $\var{appendB} :: \BX{[\con{Int}]} \to \BX{[\con{Int}]} \to \BX{[\con{Int}]}$.



The next step is to build a sketch based on the unidirectional definition given in the
specification. The type we have from above straightforwardly implies that the $\CASE$ construct
in $\var{append}$'s definition is to be replaced by the bidirectional $\BCASE$, which
expects exit conditions and reconciliation functions to be added (as holes ($\GHole$) in the sketch).
\begin{codemath}
  appendB = \lambda \var{xs}.\, \lambda \var{ys}.\, \BCASE~\var{xs}~\BOF~\{
  \begin{tarray}[t]{lll}
    [\,] \to \var{ys} ~
    \begin{tarray}[t]{ll}
      \BWITH~\GHole ~\BBY  ~\GHole;
    \end{tarray}
    \\
    (a:x) \to a : \var{appendB} \A x \A \var{ys} ~\BWITH~\GHole ~ \BBY  ~\GHole \}
  \end{tarray}
  &&\end{codemath}
Both the exit conditions and reconciliation functions are simply
unidirectional functions. Thus in theory, one can try to use a
generic synthesizer to generate them. However, this naive method will miss out
on a lot of information that we know about these functions. Recall that, given a $\BCASE$ branch
$\var{p} \to \var{e} $, its corresponding exit condition must
return true for all possible evaluations of $\var{e}$; similarly, the
results of its reconciliation function must match \var{p} and
the second argument of the reconciliation function must be an evaluation result of
\var{e}. We therefore capture such knowledge with
specialized sketches, which make use of
two types of specialized holes that are parameterized with additional information: \emph{exit-condition hole} ($\EHoleS{\var{e}}$), and \emph{reconciliation-function hole} ($\RHoleS{p}{e}$). This results in the following sketch for this example.
%
\begin{codemath}
  appendB & = \lambda \var{xs}.\, \lambda \var{ys}.\, \BCASE~\var{xs}~\BOF~\{
  \begin{tarray}[t]{lll}
    [\,] \to \var{ys} ~
    \begin{tarray}[t]{ll}
      \BWITH~\EHoleS{\var{\var{ys}}} ~\BBY  ~\RHoleS{[\,]}{\var{ys}};
    \end{tarray}
    \\
    (a:x) \to a : \var{append} \A x \A \var{ys}
    \\
    \qquad\quad \begin{tarray}[t]{lll}
      \BWITH & \EHoleS{a : \var{append} \A x \A \var{ys}} \\
      \BBY   & \RHoleS{(a:x)}{a : \var{append} \A x \A \var{ys}}\}
    \end{tarray}
  \end{tarray}
  &&\end{codemath}

In the spirit of component-based synthesis~\cite{DBLP:conf/icse/JhaGST10,DBLP:conf/popl/FengM0DR17}, we generate code by composing components from a library
that includes 
$\CASE$ and $\BCASE$ expressions,
$\con{Bool}$ constructors and operators, as well as list and tuple
constructors. As we will explain in Section~\ref{sec:synthesis-input},
this library can be augmented with auxiliary components provided by the
user.

In this example, the sketch generation is quite deterministic.
In general, especially when multiple functions must be synthesized together and auxiliary components are provided,
there could be
multiple candidate sketches. In such a case, we use a lazy approach that nondeterministically tries exploring one candidate and
generating any other.
%
%

\paragraph{{\bf Sketch completion step I: shape-restricted holes}}
With the sketch ready, we can proceed to fill the holes.
As a first step in the sketch completion process, we make use of the
information captured by the specialized holes to generate some parts
of the code for exit conditions and reconciliation functions. This step
does not involve any search.
%
\bgroup
\setlength{\jot}{0pt}
\begin{codemath}
  \var{appendB} = \lambda \var{xs}.\, \lambda \var{ys}.\, \BCASE~\var{xs}~\BOF~\{
  \begin{tarray}[t]{lll}
    [\,] \to \var{ys} ~
    \begin{tarray}[t]{ll}
      \BWITH &\lambda v.\,\CASE~v~\OF~\{ x \to \Hole;\; \dontcare \to \con{False} \}\\
      \BBY   &\lambda s. \lambda v.\,\CASE~v~\OF~\{ x \to 
      \SHole{[\,]}
      \};
    \end{tarray}
    \\
    (a:x) \to a : \var{append} \A x \A \var{ys}
    \\
    \qquad \begin{tarray}[t]{lll}
      \BWITH &\lambda v.\,\CASE~v~\OF~\{ z : \var{zs}\to \Hole;\; \dontcare \to \con{False} \}\\
      \BBY   &\lambda s. \lambda v.\,\CASE~v~\OF~\{ z : \var{zs} \to 
      \SHole{a:x}
      \}\}
    \end{tarray}
  \end{tarray}
  &&\end{codemath}
\egroup
The specialized holes are replaced with $\lambda$-abstractions with $\CASE$ structures.
The result involves a different type of holes we call \emph{shape-restricted holes} ($\SHole{p}$);
such holes can only be filled with expressions that may match the pattern $p$. For example, for $\SHole{a : x}$, the empty list is not a valid candidate. A generic hole ($\Hole$) is a special case where the pattern is a wildcard that matches every term.

%

For exit conditions, the translation used the information encoded by
exit condition holes to figure out when $\con{False}$ should be
returned---recall that, for a $\BCASE$ branch $\var{p \to
    \var{e}}$, exit conditions should return $\con{False}$ for any
results that cannot be produced by \var{e}.  In the case of
\var{appendB}, this means that for the second branch in the sketch,
the exit condition must return $\con{False}$ for any empty list.
(Here, \var{z} and \var{zs} are fresh variables.)
For the first
branch, this information does not help us eliminate any candidates.
(Again, \var{x} is a fresh variable.)
The $\CASE$ construct
generation uses all the information encoded by the exit condition
holes. Consequently, the holes left in the
sketch are generic ones.

For reconciliation functions, the newly generated
shape-restricted holes capture the fact that for a $\BCASE$ branch $\var{p} \to \var{e}$,
the result of the reconciliation function must match $p$. Thus, the first branch
of \var{appendB} has $\SHole{[\,]}$ while the second one $\SHole{a:x}$.
We further know that the second argument of the reconciliation function
must be a result of \var{e}, which allows us to generate the $\CASE$
structure shown in the sketch.


\paragraph{{\bf Sketch completion step II: search and filtering}}

The last step is to fill the remaining shape-restricted holes.
At this point, we leave off using the information in the unidirectional input program, and turn our attention to the input/output example(s).
To fill the holes,
we generate $\beta$-normal forms where functions are $\eta$-expanded, and filter the candidates
by checking against the examples(s). 
A problem with using the example(s) to filter out incorrect candidates
is that it is for the whole program, which includes several holes. A
naive use of the example(s) means that filtering has to be delayed until
late in the synthesis process when all the holes are filled.
This is inefficient.

Conversely, our ideal goal is to have a modular filtering process, where we can
simultaneously check candidate exit conditions and reconciliation functions independently of each other.
For this purpose, our solution is to leverage domain-specific knowledge of HOBiT.
\revised{
Specifically, we make use of the fact that $\PUT \A (s,v)$ and $\GET
  \A (\PUT \A (s,v))$ must follow the same execution trace in terms of
  taken branches, as explained in the discussion on round-tripping in HOBiT (see the corresponding paragraph in Section~\ref{sec:prelim}).}
This enables us to fix the control flow of the $\PUT$ behavior for the given input/output example(s) without referring to exit conditions,
so that we can separate the search for exit conditions from reconciliation functions.
We will discuss this in more
detail later in the overview, as well as in Section~\ref{sec:synthesis-search-and-filtering}.


Moreover, 
the use of the trace information also enables us to
address the issue of non-terminating $\PUT$ executions. In a naive
generate-and-test synthesis approach, some of the generated candidates may
be non-terminating, which poses issues for the testing phase. As we assume
that the $\PUT$ execution must always follow the finite branching trace of $\GET$,
we never generate such programs.
Here, we assumed that the input/output unidirectional
program is terminating for the original and updated sources of the
input/output examples. More details on this will be presented in Section~\ref{sec:synthesis-prefiltering}.

\paragraph{Filtering of exit conditions based on branch traces}

%
We continue with the partially filled sketch for \var{appendB} above (reproduced below), with the holes numbered for easy reference.
\bgroup
\setlength{\jot}{0pt}
\begin{codemath}
  \var{appendB} = \lambda \var{xs}.\, \lambda \var{ys}.\, \BCASE~\var{xs}~\BOF~\{
  \begin{tarray}[t]{lll}
    [\,] \to \var{ys} ~
    \begin{tarray}[t]{ll}
      \BWITH &\lambda v.\,\CASE~v~\OF~\{ x \to \Hole_1;\; \dontcare \to \con{False} \}\\
      \BBY   &\lambda s. \lambda v.\,\CASE~v~\OF~\{ x \to 
      \SHole{[\,]}_3
      \};
    \end{tarray}
    \\
    (a:x) \to a : \var{append} \A x \A \var{ys}
    \\
    \qquad \begin{tarray}[t]{lll}
      \BWITH &\lambda v.\,\CASE~v~\OF~\{ z : \var{zs}\to \Hole_2;\; \dontcare \to \con{False} \}\\
      \BBY   &\lambda s. \lambda v.\,\CASE~v~\OF~\{ z : \var{zs} \to 
      \SHole{a:x}_4
      \}\}
    \end{tarray}
  \end{tarray}
  &&\end{codemath}
\egroup

What are the constraints on the holes that we can derive from the input/output example below?
\[
  \texttt{:put} \A (\var{uncurryB}~ \var{appendB}) \A ([1,2,3], [4,5]) \A [6,2]  = ([6,2], [])
\]
As mentioned above, $\texttt{:put} \A (uncurryB~ appendB) \A ([1,2,3], [4,5]) \A [6,2]$ must choose the branches chosen by $\texttt{:get} \A (\var{uncurryB} \A \var{appendB}) \A ([6,2],[])$.
We shall call a history of chosen branches a \emph{branch trace}.
For $\texttt{:get} \A (\var{uncurryB} \A \var{appendB}) \A ([6,2],[])$, the branch trace is:
\begin{enumerate}
  \item[(i)] the cons branch (where $\var{xs}$ is $[6,2]$),
  \item[(ii)] the cons branch (where $\var{xs}$ is $[2]$),
  \item[(iii)] the nil branch (where $\var{xs}$ is $[]$).
\end{enumerate}
We now follow the same trace for $\texttt{:put} \A ((uncurryB~ appendB) ~([1,2,3], [4,5])~ [6,2])$ and each
step will give rise to a constraint on the exit condition of the branch.
\begin{itemize}
  \item[(i)] $a : \var{append} \A x \A \var{ys}$ (and therefore the $v$) has the value of the updated view $[6,2]$, and $\Hole_2$ must evaluate to $\con{True}$ in this context. Therefore, $\Hole_2[6/z, [2]/\var{zs}, [6,2]/v] \equiv \con{True}$.
  \item[(ii)] $a : \var{append} \A x \A \var{ys}$ (and therefore the $v$) has the value of the updated view $[2]$, and $\Hole_2$ must evaluate to $\con{True}$ in this context. Therefore, $\Hole_2[2/z, [\,]/\var{zs}, [2]/v] \equiv \con{True}$.
  \item[(iii)] $\var{ys}$ (and therefore the $v$) has the value of $[\,]$, and $\Hole_1$ must evaluate to $\con{True}$ in this context. Therefore, $\Hole_1[[\,]/x, [\,]/v] \equiv \con{True}$.
\end{itemize}
These constraints are useful in generating the exit conditions independently. As a matter of fact, in the case of $\var{appendB}$ both
$\Hole_1$ and $\Hole_2$ are simply filled by the expression $\con{True}$ which satisfies all the constraints.

There are no trace constraints generated for holes 3 and 4 though. So they will be generated according to the shape restrictions only.
Hole 3 must be $[\,]$ while Hole 4 can be filled by a non-empty list.
Recall that, in this example, the reconciliation functions of the
cons branch are never used. And therefore, arbitrary default terms will fill Hole 4 just fine, which produces the output we saw
at the beginning of this subsection.


\paragraph{Filtering of reconciliation functions based on branch traces}
The branch traces are also used to filter reconciliation functions. (This is not needed in this example
as the nil branch was already fixed in the filling of shape-restricted holes and the cons branch can be arbitrary.)
The important insight here is that reconciliation functions can be filtered independently from the exit conditions, resulting in
significant efficiency gain. The reason is that the branch traces carry all the information that is needed to test reconciliation functions (recall
that the exit conditions are only for determining branching in backward execution; and since the branching is known in the branch traces there is no need for exit conditions.). We will see examples of this in Section~\ref{sec:synthesis-prefiltering}. 

%
%


In the rest of this section, we will go through each step of the synthesis process in detail.

\subsection{Input to Our Method}
\label{sec:synthesis-input}

Remember from the overview that our technique takes as input some typed unidirectional code and a set of input/output examples illustrating the backward transformation.
Formally, this translates to the following 4-tuple $I = (P, \Gamma, f_1, \ExSet)$:
%
\begin{itemize}
  \item $P = \{f_i = e_i \}_i$ is a program in the unidirectional fragment of HOBiT, 
        where $f_i = e_i$ stands for a function/value definition of $f_i$ by the value of $e_i$.
  \item $\Gamma = \{ f_i : A_i \}_i$ is a typing environment for $P$; \ie,
        each $e_i$ has type $A_i$ under $\Gamma$. 
  \item $f_1$ is the entry point function, whose type is expected to have the form $\sigma_1 \to \tau_1$;%
        \footnote{We use metavariables $A, B,\dots$ for types in general and $\sigma, \tau,\dots$ for those that can be sources or views. In \citet{Matsuda&Wang:2018:HOBiT}, the latter kind of types do not contain $\BX{}$ and $\to$, but their implementation does not distinguish the two (which in fact is safe). Thus, we do not strictly respect the restriction on $\sigma$-types in our technical development. 
        }
        this is used to prune the search space as explained in Section~\ref{sec:template-generation}. 
      \item $\ExSet = \{ (s_k, v_k, s_k') \}_k$ is a (finite) set of well-typed input/output examples for a bidirectional version of the entry point $f_1$;
        for the $k$-th example, $s_k$ is the original source, $v_k$ is the updated view and $s_k'$ is the updated source.
\end{itemize}

The input program $P$ may contain functions that are not
reachable from the entry point but can be used during program generation.
We call such functions \emph{auxiliary functions} and add them to our
library of default synthesis components.  As mentioned earlier in
Section~\ref{sec:overview}, the default library includes $\CASE$ and
$\BCASE$ expressions, $\con{Bool}$ constructors and operators, as well
as list and tuple constructors.



\begin{example}[$\var{append}$]
  \label{example:append}
  For the \var{appendB} example, the input is formally expressed as: 
  \begin{align*}
    P_\mathrm{app}      & = \{ \var{appendB} = \lambda \var{xs}.\, \lambda \var{ys}.\, \CASE~\var{xs}~\OF~\{ [\,] \to \var{ys} ;\; (a:x) \to a : \var{appendB} \A x \A \var{ys}, \var{uncurryB} = \dots \} \} \\
    \Gamma_\mathrm{app} & = \{ \var{appendB} : [\con{Int}] \to [\con{Int}] \to [\con{Int}], \var{uncurryB} : \dots \}                                                                                         \\
    f_{1\mathrm{app}}   & = \var{uncurryB} \A \var{appendB}                                                                                                                                                   \\
    \ExSet_\mathrm{app} & = \{ (([1,2,3], [4,5]), [6,2], ([6,2],[])) \}\text{.}                                                                                                                               
  \end{align*}
  Here, we omit the definition and the type of $\var{uncurryB}$ but state it is a part of the input program. \qed
\end{example}

\subsection{Generation of Sketches}
\label{sec:template-generation}
\label{sec:synthesis-template-generation}

As shown in the overview, we start by generating bidirectional sketches from the unidirectional code.
The basic idea of the sketch generation is to replace unidirectional constructs with bidirectional ones nondeterministically:
when $\CASE$ is replaced with $\BCASE$, exit conditions and reconciliation functions are left as holes.
Interestingly, replacing all unidirectional constructs (if they have corresponding bidirectional ones) may not be the best solution;
as demonstrated in $\var{appendBc} :: \BX{[a]} \to [a] \to \BX{[a]}$ in Section~\ref{sec:prelim},
we sometimes need to leave some parts unidirectional to achieve the given bidirectional behavior.


The starting point of sketch generation is deciding the type of the
target function.  We expect the unidirectional code to
contain an entry point function $f_1 : \sigma_1 \to \tau_1$ (e.g., \var{uncurry~append} in Example~\ref{example:append}). This helps
us reduce the number of generated type signatures as we know that the
target entry point function to be synthesized must have type
$\BX{\sigma_1} \to \BX{\tau_1}$. Also, we further prune the search space
by eliminating type signatures that do not obey the call context in the
input/output examples.

\paragraph{Type signature generation}
We first define the relation $A \leadsto A'$ as:
$A'$ is the type obtained from $A$ by replacing an arbitrary number of sub-components $\sigma$ in $A$ by $\BX{\sigma}$ nondeterministically,
as long as $\sigma$ does not contain function types.
We do not replace $\sigma$ containing function types to avoid generating apparently non-useful types
such as $\BX{(\con{Int} \to \con{Int})}$ and $\BX{[\con{Int} \to \con{Int}]}$.
Next, we provide the typing environment generation relation $\Gamma \leadsto \Gamma'$, where $\Gamma'$ is the typing environment corresponding to the bidirectional program.
\begin{definition}[Generation of Typing Environment]
  For $\Gamma = \{f_1 : \sigma_1 \to \tau_1 \} \cup \{f_i : A_i\}_{i > 0}$, the typing environment generation relation $\Gamma \leadsto \Gamma'$ is defined if
  $\Gamma' = \{ f_1 : \BX{\sigma_1} \to \BX{\tau_1} \} \cup \{ f_i : A_i' \}_{i > 0}$,
  where $A_i \leadsto A_i'$ for each $i > 0$. \qed
\end{definition}

\paragraph{Type-directed sketch generation}

Once we have (a candidate) typing environment $\Gamma'$, the next step is to generate corresponding sketches in a type-directed manner.
Very briefly, the type system in HOBiT~\cite{Matsuda&Wang:2018:HOBiT} uses a dual context system~\cite{DaviesPf01}. The typing relation can be written as $\Gamma ; \Delta \vdash e : A$,
where $\Gamma$ and $\Delta$ respectively are called unidirectional and bidirectional typing environments, and hold
variables introduced by unidirectional and bidirectional contexts respectively.
\ifFullVersion
See Appendix~\ref{sec:hobit-typing-complete} for the concrete typing rules.
\fi

The sketch generation is done by using a relation $\GenTmp{\Gamma'}{\Delta'}{A'}{\Gamma}{e}{A}{e'}$, which
reads that, from a term-in-context $\Gamma ; \emptyset \vdash e : A$, sketch $e'$ is generated according to the given target typing environments $\Gamma'$ and $\Delta'$, and target type $A'$
so that $\Gamma' ; \Delta' \vdash e' : A'$ holds after the sketch has been completed (i.e., no unfilled holes) in a type-preserving way. 
Notice that $\Gamma'$, $\Delta'$ and $A'$ are also a part of the input in $\GenTmp{\Gamma'}{\Delta'}{A'}{\Gamma}{e}{A}{e'}$; \ie, its outcome is only $e'$.
\ifFullVersion
We omit the concrete generation rules in the main text but put them in Fig.~\ref{fig:gen-template} in Appendix~\ref{sec:template-generation-rules-full}, because they are technical but straightforward.
\fi

We note that the rules are overlapping (i.e., several may be applicable at a given step), which makes sketch generation nondeterministic. 
%
The sketch generation is defined formally as below.
\begin{definition}[Type-Directed Sketch Generation]
  Suppose that $\Gamma \leadsto \Gamma'$.
  Then, for $P = \{ f_i : e_i \}_i$,
  the sketch generation relation $P \leadsto P'$ is defined if
  $ P' = \{ f_i : e'_i \}_i$,
  where $\GenTmp{\Gamma'}{\emptyset}{\Gamma'(f_i)}{\Gamma}{e_i}{\Gamma(f_i)}{e'_i}$. \qed
\end{definition}



\subsection{Sketch Completion Step I: Shape-Restricted Holes}
\label{sec:filling-holes}

In general, there will be several possible sketches for a given
unidirectional program.
As mentioned in Section~\ref{sec:overview}, in such a case, 
we use a lazy approach that nondeterministically tries exploring one candidate and generating any other.
%
In this section we describe 
the sketch exploration
process. In particular, we start by using the information captured by
the specialized holes to generate parts of the code for exit
conditions and reconciliation functions. 
\subsubsection{Handling Exit Condition Holes}
\label{sec:synthesis-with}


Remember that an exit condition matching a hole $\EHoleS{\var{e}}$ should return False for any results that cannot be produced by \var{e}.
Then, our idea here is to generate code that returns $\con{False}$ for values that are obviously not the result of $e$.
For example, for $\EHoleS{a : \var{append} \A \var{x} \A \var{ys}}$,
we generate code returning $\con{False}$ for the empty list.

Let us write $\mathcal{P}(e)$ for a pattern that represents an obvious shape of $e$, defined as follows (where $\con{C}$ is a constructor):
\begin{revisedblock}
\[
\mathcal{P}(e) = \begin{cases}
  \con{C} \A \mathcal{P}(e_1) \A \dots \A \mathcal{P}(e_n) & \text{if}~e = \con{C} \A e_1 \A \dots \A e_n \\
  x & \text{otherwise ($x$ : fresh)}
 \end{cases}
\]
\end{revisedblock}
For example, we have $\mathcal{P}(a : \var{append} \A \var{x} \A \var{ys}) = \revised{\mathcal{P}(a) : \mathcal{P}(append ~x ~ys)} = z : \var{zs}$, where $z$ and $\var{zs}$ are fresh,
\revised{conforming to the second case above}.
It is quite apparent that any result of $e$ matches with $\mathcal{P}(e)$; in other words, values that do not match with $\mathcal{P}(e)$ cannot be
a result of $e$.
Using $\mathcal{P}(e)$, we concretize exit-condition holes as below.
\begin{definition}[Partial completion of exit condition holes]
  Let $p_e$ be a pattern $\mathcal{P}(e)$. Then, the exit-condition-hole partial completion relation
  $\EHoleS{e} \leadsto e'$, which reads hole $\EHoleS{e}$ is filled by $e'$, is defined by the rule
  \[
    \pushQED{\qed}
    \EHoleS{e} \leadsto \lambda s.\,\CASE~s~\OF~\{ p_e \to \Hole ;\; \dontcare \to \con{False} \} \qedhere
    \popQED
  \]
\end{definition}

Note that the resulting sketch will contain a generic hole $\Hole$, whose shape is no longer constrained.
\revised{For example, $\EHoleS{a : \var{append} \A \var{x} \A \var{ys}}$ is converted as follows}
\[
  \revised{\EHoleS{a : \var{append} \A \var{x} \A \var{ys}} \leadsto \lambda s.\,\CASE~s~\OF~\{ \var{z} : \var{zs} \to \Hole ;\; \dontcare \to \con{False} \} }
\]

\subsubsection{Handling Reconciliation Function Holes}

Remember that the role of a reconciliation function associated with a branch is to reconcile the original source with the branch
by producing a new ``original source'' matching the branch (Section~\ref{sec:prelim}).
Thus, when the branch has the form $p \to e$, the reconciliation function must return a value of the form $p[\V{v/x}]$ where $\{\V{x}\} = \fv(p)$.
Hence, a natural approach is to generate reconciliation functions of the form
$\lambda s. \lambda \mathcal{P}(e). p[\V{e/x}]$.

However, only considering expressions of the aforementioned form limits the use
of user-specified auxiliary functions in reconciliation functions.
%
Instead, we generate reconciliation functions of the form
$\lambda s. \lambda \mathcal{P}(e). \SHole{p}$.
Recall that the shape-restricted hole $\SHole{p}$ will be filled by expressions shaped $p$.
This idea is formally written as below.
\begin{definition}[Partial completion of reconciliation function holes]
  Let $p_e$ be a pattern $\mathcal{P}(e)$. Then, the partial completion relation for the reconciliation function hole,
  $\RHoleS{p}{e} \leadsto e'$, which reads hole $\RHoleS{p}{e}$ is filled by $e'$, is defined by the rule
  \[
    \pushQED{\qed}
    \RHoleS{p}{e} \leadsto
    \lambda s. \lambda v.\,
    \CASE~v~\OF~\{ p_e \to \SHole{p} \} \qedhere
    \popQED
  \]
\end{definition}

\subsection{Sketch Completion Step II: Search and Filtering}\label{sec:synthesis-search-and-filtering}
The last step is to fill the remaining shape-restricted and generic
holes.  This process involves type-directed generation of candidates and filtering based on user-provided input/output examples.
For simplicity of presentation, we do not explicitly capture the type of the code to be
generated in the shape-restricted holes; instead, we recover it
from the sketch and typing environment $\Gamma$.

\subsubsection{Generating Candidates for Shape Restricted Holes}
\label{sec:synthesis-filling-generic-holes}
In this section, we describe the process of filling in shape restricted holes $\SHole{p}$.
To achieve this, we generate terms of shape \var{p} in $\beta$-normal forms where functions are $\eta$-expanded.
Specifically, we produce expressions $U^p$ in the following grammar, which restricts their shape to \var{p}: 
\[
  \begin{tarray}{l@{\quad}l}
    \begin{tarray}{lcll}
      U^p &::=& V^p \mid \CASE~x \A V_1 \A \dots \A V_n~\OF~\{ p_i \to U_i^p \}_i
    \end{tarray}\\
    \begin{tarray}{lcl@{\qquad}l}
      V^p &::=& \lambda x. U & (p = x) \\
      &\mid& x \A V_1 \A \dots \A V_n \\
      &\mid& \con{C} \A V_1^{p_1} \A \dots \A V_n^{p_n} & (p = \con{C} \A p_1 \A \dots \A p_n \text{~or \(p,p_1,\dots,p_n\) are all variables})
    \end{tarray}
  \end{tarray}
\]
To simplify the presentation, we omit $p$ and write $V$ or $U$ if $p$ is a variable.
In this grammar, the purpose of $U$ is to have $\CASE$s in the outermost positions (but inside $\lambda$);
a $\CASE$ in a context $K[\CASE~e~\OF~\{ p_i \to e_i \}_i]$
can be hoisted as $\CASE~e~\OF~\{ p_i \to K[e_i] \}_i$, which is a transformation known as commuting conversion.
If $p$ is not a variable, we can only generate constructors as specified by $p$
($p = \con{C} \A p_1 \A \dots \A p_n $  or  $(p,p_1,\dots,p_n)$ are all variables).
Otherwise, if $p$ is a variable, the only knowledge we assume about it is its type. Thus, any of the
productions for $V^{p}$ would be considered.
Note that $x$/$C$ are drawn from the current context; \ie, they may be components provided by users.


Types are used for two purposes in this type-directed generation.
The rather obvious purpose is to limit the search space for $x$ and $\con{C}$; notice that, since we know their types,
we also know the types of their arguments allowing us to perform type-directed synthesis for them as well.
The other purpose is to reduce redundancy with respect to $\eta$-equivalence by only generating
$\lambda x.U$ for function types and $\CASE$s only for non-function types.
A caveat is the generation of $x \A V_1 \A \dots \A V_n$ at the scrutinee position of $\CASE$, which cannot be done in a type-directed way as its type is not given \textit{a priori};
instead, its type is synthesized by using the type of $x$.


\subsubsection{Filtering Based on Branch Traces}
\label{sec:synthesis-prefiltering}

\revised{As explained in the discussion on round-tripping in HOBiT (see the corresponding paragraph in Section~\ref{sec:prelim}), we leverage
the fact that $\PUT \A (s,v)$ and $\GET
  \A (\PUT \A (s,v))$ must follow the same execution trace in terms of
taken branches.} 
%
This enables us to fix the control flow of the $\PUT$ behavior for the
given input/output example(s) without referring to exit conditions, making it possible
to separate dependent synthesis tasks.

While the example \var{appendB} in Section~\ref{sec:overview} only
showed how filtering works for exit conditions, we provide next one example
illustrating filtering based on branch traces
for both exit conditions and reconciliation functions. We conclude
with a discussion on pruning away programs that would otherwise cause
non-terminating $\PUT$ executions.
\ifFullVersion
\revised{We shall refer interested readers to the extended version~\cite{extended-version} of this paper for the formal description of this trace-based filtering process.}
\fi

\paragraph{Example of filtering exit conditions and reconciliation functions based on branch traces}

Consider the following  program
\begin{codemath}\bb
  f :: \con{Either} \A \con{Int} \A \con{Int} \to \con{Bool} \\
  f \A x = \CASE~x~\OF~\{ \con{Left}~x \to \con{True};\; \con{Right}~x \to \con{False} \}
  \ee&&\end{codemath}
that comes with two input/output examples that negate the view and cause the sources to flip: $\ExSet = \{(\con{Left} \A 42, \con{False},\con{Right} \A 42),~ (\con{Right} \A 42, \con{True},\con{Left} \A 42)\}$.
Suppose that from the given unidirectional code, we obtain the following candidate before any filtering is done.
\begin{codemath}\bb
  f \A x = \BCASE~x~\BOF~\{
  \bbt \con{Left} \A x \to \bxcon{True}
  &\BWITH& \lambda v. \CASE~v~\OF~\{ \con{True} \to \Box_1;\; \con{False} \to \con{False} \} \\
  &\BBY~ & \lambda s.\lambda v. \CASE~v~\OF~\{ \con{True} \to \Box_2 \} ;
  \\
  \con{Right} \A x \to \bxcon{False} \,
  &\BWITH&\lambda v. \CASE~v~\OF~\{ \con{False} \to \Box_3;\; \con{True} \to \con{False} \} \\
  &\BBY~ &\lambda s.\lambda v. \CASE~v~\OF~\{ \con{False} \to \Box_4 \} \}
  \ee
  \ee&&\end{codemath}

We first discuss filtering of exit conditions.
For the first example, $\texttt{:get}~\var{f} ~(\con{Right}~42)$ takes the second branch, meaning that we obtain the constraint
$\Box_3[\con{False}/v] \equiv \con{True}$.
For the second example, $\texttt{:get}~\var{f} ~(\con{Left}~42)$ takes the first branch, generating the constraint $\Box_1[\con{True}/v] \equiv \con{True}$.
A solution for these constraints is $\Box_1 = \con{True}$ and $\Box_3 = \con{True}$.


Now, let us focus on the reconciliation functions.
If we consider the branch trace generated by the $\GET$ and evaluate the $\PUT$ for the given examples, we obtain
the following constraints:
\begin{itemize}
  \item  For $\texttt{:put}~\var{f}~ (\con{Left} \A 42)~ \con{False}$, we must switch branches to the second branch, meaning that the
        reconciliation function corresponding to the second branch gets triggered, generating the constraint:
        $\texttt{:put}~\var{f} (\Hole_4[\con{False}/v])~ \con{False} = \con{Right}~42$.

  \item For $\texttt{:put}~\var{f}~ (\con{Right} \A 42)~ \con{True}$, we must switch branches to the first branch, meaning that the
        reconciliation function corresponding to this branch gets triggered, generating the constraint:
        $\texttt{:put}~\var{f} (\Hole_2[\con{True}/v])~ \con{True} = \con{Left}~42$.
\end{itemize}
From these constraints, one possible solution is $\Box_2
  = \con{Left} \A 42$ and $\Box_3 = \con{Right} \A 42$.  While this
solution obeys the given example, a better one would be $\Box_2
  = \CASE~s~\OF~\{ \con{Left} \A x \to s; \con{Right} \A y \to \con{Left} \A y \}$ and $\Box_3 = \CASE~s~\OF~\{ \con{Left} \A x \to \con{Right} \A x; \con{Right} \A y \to s\}$; each $s$ in the branch bodies in $\Box_2$ and $\Box_3$ can be arbitrary, as they will never be used.
The suboptimal solution could be filtered out by
our synthesis engine if other examples such as
$\texttt{:put}~\var{f}~(\con{Left} \A 37, \con{False}) = \con{Right} \A 37$
were provided by the user.

As a note, for both the previous example and the running example $\var{appendB}$ in Section~\ref{sec:overview},
we only generate positive constraints (i.e., that evaluate to $\con{True}$) for the holes in exit conditions.
However, in certain cases, negative constraints (i.e., that evaluate to $\con{False}$) may also be generated.
This happens when branch switching implies that the original branch's exit condition evaluates to $\con{False}$.
We encountered such situations for \var{lengthTail} and \var{reverse} in Section~\ref{sec:experiments}.
In such a case, the choice of reconciliation functions may affect the generated constraints as they specify new sources.




\paragraph{Discussion on pruning non-terminating programs based on branch traces}
Using branch traces also helps us prune away programs that would cause
non-terminating $\PUT$ executions.  It is known that synthesis of recursive functions is a
challenging problem~\cite{DBLP:conf/cav/AlbarghouthiGK13}, especially
for programming-by-examples, because a synthesized function may
diverge for a given example.  Waiting for a timeout is inefficient
and there is no clear way to set an appropriate time limit.
In our approach, assuming that the given $\GET$ execution is terminating for the
input/output examples, we never generate such diverging candidate programs.
The reason is that we only generate programs whose $\PUT$ execution follows the
finite branch trace of the $\GET$, and are thus terminating.

\subsection{Heuristics}
\label{sec:heuristics}
%
%
In this section, we discuss some heuristics we found effective when exploring the search space.

%

\emph{Assigning costs to choices.}
Our generation is prioritized by assigning a positive cost to each
nondeterministic choice in the sketch generation.
Programs with lower costs are generated earlier than those with higher costs.  An advantage of
this approach is that it is easy to integrate with lazy nondeterministic generation
methods~\cite{DBLP:journals/jfp/FischerKS11}, which is the core of our
prototype implementation.  Another advantage is that smaller programs
naturally have higher priority (\ie, lower costs), as the generation of
large programs usually involves many choices, reflecting our belief
that smaller programs are typically preferable.

\emph{Canonical forms of $\con{Bool}$-typed expressions.}
Generation of $\con{Bool}$-typed expressions is a common task, especially in our context as
exit conditions always return $\con{Bool}$ values. However, a naive generation of $\con{Bool}$-typed expressions may lead to redundancies, for example, $\con{True} \mathbin{\texttt{\&\&}} e$ and $e$ may be considered two distinct expressions during the search.
So when filling holes of type $\con{Bool}$, we generate expressions in disjunctive normal form, in which
atomic propositions are expressions of the form $x \A V_1 \A \dots \A V_n$ with $x : A_1 \to \dots \to A_n \to \con{Bool} \in \Gamma$. 
While this eliminates redundancy due to distributivity, associativity and zero and unit elements,
it does not address commutativity and idempotence.
Provided that there is a strict 
total order $\prec$ on expressions, both sorts of redundancy could be addressed easily by
generating $e_2$ \emph{after} $e_1$ so that $e_1 \prec e_2$ holds.
Currently we do not do this in our implementation in order to avoid
the additional overhead of checking $e_1 \prec e_2$.



\emph{Other effective improvements.}
In addition to the heuristics mentioned above, we make use of some simple but effective techniques.
For example, for $\BCASE$ with a single branch, we do not try to synthesize
exit conditions or reconciliation functions. An exit condition $\lambda \dontcare.\con{True}$ and a reconciliation function $\lambda s.\lambda v.s$ suffice for such a branch.
We do not generate redundant case expressions such as
$\lambda s.\lambda v.\CASE~ v~ \OF \{ z \to \cdots \}$. 
When the pattern $p$ of a branch does not contain any variables, we deterministically choose $\lambda \dontcare.\lambda \dontcare.p$ as its reconciliation function.
For a $\CASE$ whose patterns $\{\mathcal{P}(e_i)\}_i$ do not overlap, we do not leave holes in exit conditions as replacing them with $\con{True}$ is sufficient.




\subsection{Soundness and Incompleteness}
\label{sec:soundness-incompleteness}

Our proposed method is sound for the given input/output examples in the sense that it synthesizes a bidirectional transformation such that 
its $\PUT$ behavior is consistent with the input/output examples, and its $\GET$ behavior coincides with the given $\GET$ program for the sources that appear in the examples. 
This is obvious because we check the conditions in the last step (\ie, filtering) in our synthesis. 
It is worth noting that the $\GET$ behavior of a synthesized function may be less defined than a given $\GET$ program, 
because our method may synthesize exit conditions that are not postconditions; recall that they are checked dynamically in HOBiT (Section~\ref{sec:prelim}). 
We heuristically try to avoid this by prioritizing $\con{True}$ over $\con{False}$ in the synthesis of exit conditions, which works effectively for all the cases discussed in Section~\ref{sec:experiments} but is not a guarantee, especially with components. 
We could address this by inferring postconditions and using them as exit conditions, which is left for future work.

In contrast, our proposed method is incomplete. This is due to the use of the sketches obtained from the unidirectional code to prune the search space. 
While this makes our approach efficient, it may remove potential solutions. Such situations are captured by the examples $\var{lines}$ and $\var{lookup}$, where the solutions do not follow the sketches,
in the experimental evaluation in Section~\ref{sec:experiment-categorize}.


\section{Experiments}\label{sec:experiments}
We implemented the proposed idea as a proof-of-concept system, {\SysName}, in Haskell\footnote{The implementation is available in the artifact \url{https://doi.org/10.5281/zenodo.5494504} or in \url{https://github.com/masaomi-yamaguchi/synbit}}.
{\SysName} is given as an extension to the original HOBiT
implementation~\cite{Matsuda&Wang:2018:HOBiT}.

We measure the effectiveness of our proposed method in the following three experiments.
\begin{itemize}
  \item Microbenchmarks, classified in terms of information loss (Section~\ref{sec:experiment-categorize}).
  \item More realistic problems including XML transformations and string parsing (Section \ref{sec:experiment-complex}).
  \item Comparisons with the other state-of-the-art synthesis methods (Section~\ref{sec:experiment-comparative}).
\end{itemize}

The experiments were conducted on a Windows Subsystem for Linux (WSL)
2 running on a laptop PC with 2.30 GHz Intel(R) Core(TM) i7-4712HQ CPU
and 16 GB memory, 13 GB out of which were assigned to WSL 2.  The host
OS was Windows 10 (build NO. 19042.685), and the guest OS was Ubuntu
20.04.1 LTS.  We used GHC 8.6.5 to compile {\SysName} with the
optimization flag \texttt{-O2}.  Execution times were measured by
Criterion\footnote{\url{https://hackage.haskell.org/package/criterion}},
a popular library in Haskell for benchmarking, which estimates the
true execution time by the least-squares method.
Any case running longer than 10 minutes was reported as a timeout.


\subsection{Microbenchmarks Classified by Information-Loss}\label{sec:experiment-categorize}

\newcommand{\NotInduction}{{-}}
\newcommand{\AllPreserved}{1}
\newcommand{\SomeLost}{2}
\newcommand{\AllLost}{3}

To construct the microbenchmarks, we classify programming problems according to the level of difficulty.
Recall that the main challenge of BX is to incorporate the information that is in the source but absent in the view in order to create an updated source.
For structure rich data represented by algebraic datatypes, this includes the structure of the source data, especially the part that the $\GET$ function recurs on. With that, we arrive at the following classes.
\begin{description}
  \item[Class \AllPreserved]
        All information of the recursion structure is present in the view (\eg, $\var{map}$).
  \item[Class \SomeLost]
        Some information of the recursion structure is present in the view (\eg, $\var{append}$).
  \item[Class \AllLost]
        No information of the recursion structure is present in the view (\eg, $\var{lookup}$).
\end{description}
The rule of thumb is that the more information is present in the view, the easier is it to define a $\PUT$ that
handles structural changes to the view. Take $\var{map}$ as an example, the function is bijective in terms of the list structure.
As a result, a $\PUT$ function can share the recursion structure of the $\GET$, mapping whatever structural changes from the view back to the source.
This becomes harder with the loss of structure information in the view. Take $\var{append}$ as an example. The boundary between the first source list, which $\GET$ recurs on, and the second source list is gone in the view. As a result, if a $\PUT$ function is to share the recursive structure of the $\GET$, the backward execution will always try to replenish the first source list first before leaving the remaining view elements as the second source list\footnote{unless we know that the second list is fixed as in $\var{appendBc}$}. This is what $\var{appendB}$ does. Any divergence from this behavior will require a different recursive
structure for $\PUT$, which drastically increases the search space as it loses the guidance of the $\GET$-based sketch.

We thus expect that the performance of \SysName~ varies according to the difficulty classes. For Class-\AllPreserved~ problems,
synthesis is likely to be successful for any given input/output examples (thus handling any structural changes);
for Class-\SomeLost~ problems, synthesis is likely to be successful for some given input/output examples;
and for Class-\AllLost~ problems, synthesis is only possible for input/output examples that are free from structural changes.

\begin{table}[t]
  \caption{The results of experiments for categorized examples}
  \label{table:result-experiments}\small
  \begin{tabular}{l|lccr}
    Problem                  & Recursion on          & Class         & Result  & Time (s)              \\\hline
    $\var{double}$           & Nat                   & \AllPreserved & Yes     & 0.050\phantom{}       \\ 
    $\var{uncurryReplicate}$ & Nat                   & \AllPreserved & Yes     & 0.050\phantom{}       \\ 
    $\var{mapNot}$           & List                  & \AllPreserved & Yes     & 0.052                 \\ 
    $\var{mapReplicate}$     & Nat and List          & \AllPreserved & Yes     & 0.13\phantom{0}       \\ 
    $\var{snoc}$             & List                  & \AllPreserved & Yes     & 0.15\phantom{0}       \\ 
    $\var{length}$           & List                  & \AllPreserved & Yes     & 0.040\phantom{}       \\ 
    $\var{lengthTail}$       & List (tail recursive) & \AllPreserved & Yes     & 0.22\phantom{0}       \\ 
    $\var{reverse}$          & List (tail recursive) & \AllPreserved & Yes     & 1.3\phantom{00}       \\ 
    $\var{mapFst}$           & List                  & \AllPreserved & Yes     & 0.016\phantom{}       \\ 
    $\var{add}$              & Nat                   & \SomeLost     & Yes     & 0.045\phantom{}       \\ 
    $\var{append}$           & List                  & \SomeLost     & Yes     & 0.034\phantom{}       \\ 
    $\var{appendBc}$         & List                  & \SomeLost     & Yes     & 5.6\phantom{00}       \\
    $\var{professor}$        & List                  & \SomeLost     & Yes     & 0.023\phantom{}       \\ 
    $\var{lines}$            & List                  & \SomeLost     & Timeout & \multicolumn{1}{c}{-} \\
    $\var{lookup}$           & List                  & \AllLost      & Timeout & \multicolumn{1}{c}{-} \\
  \end{tabular}
\end{table}
The benchmark programs and the synthesis results are summarized in Table~\ref{table:result-experiments}, which should be
read together with Table~\ref{table:examples-experiments} where
the input/output examples used for the experiments are shown (which can also be used as a reference for the forward execution behaviors of the
input functions).
Also, we used the following auxiliary functions: equality over natural numbers for \var{lengthTail},
\var{reverse} and \var{appendBc}, and \var{length} in addition for the latter two.
The definitions of all the functions listed and the full synthesis results can be found in the artifact \footnote{\url{https://doi.org/10.5281/zenodo.5494504}} or the repository\footnote{\url{https://github.com/masaomi-yamaguchi/synbit}} together with the  implementation.

\begin{table}
  \caption{
    Input/output examples: for readability, we shall write $n$ for $\con{S}^n \A \con{Z}$ (integer constants are also used in $\var{snoc}$, $\var{reverse}$, $\var{mapFst}$ and $\var{append}$), and
    $\var{st}_n$/$\var{pr}_n$/$\var{pr}_n'$ for $\con{Student} \A \texttt{"st\(n\)"}$/$\con{Professor} \A \texttt{"pr\(n\)"}$/$\con{Professor} \A \texttt{"pr\(n\)'"}$.
  }
  \label{table:examples-experiments}
  \footnotesize
  \setlength{\tabcolsep}{4pt}
  \begin{tabular}{@{}l|llll@{}}
    Program
                        & Original Source                                                 & (Original View)                                                                & Updated View                                    & Updated Source                                              \\\hline
    \multirow{2}{*}{$\var{double}$}
                        & 1                                                               & 2                                                                              & 6                                               & 3                                                           \\
                        & 5                                                               & 10                                                                             & 4                                               & 2
    \\\hline
    \multirow{2}{*}{$\var{uncurryReplicate}$}
                        & (\texttt{'a'}, 2)                                               & \texttt{"aa"}                                                                  & \texttt{"bbb"},                                 & (\texttt{'b'}, 3)                                           \\
                        & (\con{True}, 3)                                                 & [\con{True}, \con{True}, \con{True}]                                           & [\con{False},\con{False}]                       & (\con{False}, 2)
    \\\hline
    \multirow{1}{*}{$\var{mapNot}$}
                        & $[\con{True}, \con{False}]$                                     & $[\con{False}, \con{True}]$                                                    & $[\con{True},\con{True},\con{False}]$           & $[\con{False},\con{False}, \con{True}]$
    \\\hline
    \multirow{2}{*}{$\var{mapReplicate}$}
                        & $[(\texttt{'b'}, 2)]$                                           & $[\texttt{"bb"}]$                                                              & $[\texttt{"aaa"}, \texttt{"b"}, \texttt{"cc"}]$ & $[(\texttt{'a'}, 3), (\texttt{'b'}, 1), (\texttt{'c'}, 2)]$ \\
                        & $[(\texttt{'a'}, 3), (\texttt{'b'}, 1), (\texttt{'c'}, 2)]$     & $[\texttt{"aaa"}, \texttt{"b"}, \texttt{"cc"}]$                                & $[\texttt{"bb"}]$                               & $[(\texttt{'b'}, 2)]$
    \\\hline
    \multirow{2}{*}{$\var{snoc}$}
                        & $([1,2,3],4)$                                                   & $[1,2,3,4]$                                                                    & $[1,2,3]$                                       & $([1,2], 3)$                                                \\
                        & $([1,2,3],4)$                                                   & $[1,2,3,4]$                                                                    & $[1,2,3,4,5,6]$                                 & $([1,2,3,4,5], 6)$
    \\\hline
    $\var{length}$
                        & $[1,2]$                                                         & $2$                                                                            & $4$                                             & $[1,2,0,0]$                                                 \\
    $/\var{lengthTail}$ & $[2,0]$                                                         & $2$                                                                            & $1$                                             & $[2]$
    \\\hline
    \multirow{2}{*}{$\var{reverse}$}
                        & $[\con{True}, \con{True}]$                                      & $[\con{True}, \con{True}]$                                                     & $[\con{False}, \con{True}, \con{True}]$         & $[\con{True}, \con{True}, \con{False}]$                     \\
                        & $[1,2,3,4]$                                                     & $[4,3,2,1]$                                                                    & $[6,5]$                                         & $[5, 6 ]$
    \\\hline
    \multirow{2}{*}{$\var{mapFst}$}
                        & $[(1,\texttt{'a'}),(2, \texttt{'b'}), (3,\texttt{'c'})]$        & $[1,2,3]$                                                                      & $[1,3]$                                         & $[(1,\texttt{'a'}), (3,\texttt{'b'})]$
    \\
                        & $[(2, \texttt{'b'}), (3,\texttt{'c'})]$                         & $[2,3]$                                                                        & $[0,1,2,3]$                                     & $[\bbt (0,\texttt{'b'}),(1,\texttt{'c'}),                   \\(2,\texttt{'a'}), (3,\texttt{'a'})]\ee$
    \\ \hline
    \multirow{2}{*}{$\var{add}$}
                        & $(2,3)$                                                         & $5$                                                                            & $7$                                             & $(2,5)$                                                     \\
                        & $(2,3)$                                                         & $5$                                                                            & $1$                                             & $(1,0)$
    \\\hline
    \multirow{2}{*}{$\var{append}$}
                        & $([1,2,3,4], [5])$                                              & $[1,2,3,4,5]$                                                                  & $[6,2]$                                         & $([6,2], [])$                                               \\
                        & $([1,2,3,4], [5])$                                              & $[1,2,3,4,5]$                                                                  & $[1,2,3,4,5,6]$                                 & $([1,2,3,4], [5,6])$
    \\\hline
    \multirow{2}{*}{$\var{appendBc}$}
                        & $\texttt{"apple"}$                                              & $\texttt{"apple;;"}$                                                           & $\texttt{"pineapple;;"}$                        & $\texttt{"pineapple"}$                                      \\
                        & $\texttt{"apple"}$                                              & $\texttt{"apple;;"}$                                                           & $\texttt{"plum;;"}$                             & $\texttt{"plum"}$
    \\\hline
    \multirow{2}{*}{$\var{professor}$}
                        & $[ \var{st}_1, \var{st}_2, \var{pr}_1, \var{st}_3, \var{pr}_2]$ & $[\var{pr}_1, \var{pr}_2]$
                        & $[\var{pr}_1', \var{pr}_2', \var{pr}_3']$                       & $[ \var{st}_1, \var{st}_2, \var{pr}_1', \var{st}_3, \var{pr}_2', \var{pr}_3']$                                                                                                                 \\
                        & $[ \var{st}_1, \var{st}_2, \var{pr}_1, \var{st}_3, \var{pr}_2]$ & $[\var{pr}_1, \var{pr}_2]$
                        & $[\var{pr}_1']$                                                 & $[ \var{st}_1, \var{st}_2, \var{pr}_1', \var{st}_3]$
    \\\hline
    \multirow{3}{*}{$\var{lines}$}
                        & $\texttt{"aa\textbackslash nbb\textbackslash ncc"}$             & $[\mathtt{"aa"}, \mathtt{"bb"},\mathtt{"cc"}]$                                 & $[\texttt{"aa"}, \texttt{"bb"}]$                & $\texttt{"aa\textbackslash nbb"}$                           \\
                        & $\texttt{"aa"}$                                                 & $[\mathtt{"aa"}]$                                                              & $[\texttt{"aa"}, \texttt{"bb"}]$                & $\texttt{"aa\textbackslash nbb"}$                           \\
                        & $\texttt{"aa\textbackslash n"}$                                 & $[\mathtt{"aa"}]$                                                              & $[\texttt{"aa"}, \texttt{"bb"}]$                & $\texttt{"aa\textbackslash nbb\textbackslash n"}$
    \\\hline
    \multirow{2}{*}{$\var{lookup}$}
                        & $([(1,10), (2,200), (3,33)], 2)$                                & $200$                                                                          & $10$                                            & $([(1,10), (2,200), (3,33)], 1)$                            \\
                        & $([(1,10), (2,200), (3,33)], 2)$                                & $200$                                                                          & $33$                                            & $([(1,10), (2,200), (3,33)], 3)$                            \\
  \end{tabular}
\end{table}

\paragraph{Class \AllPreserved} As we can see, \SysName~handles programs in this class with ease.
An interesting case is $\var{reverse}$. On the conceptual level, the function is embarrassingly bijective
and should be straightforward to invert. However, in practice the story is much more complicated, especially for the
linear-time accumulative list reversal (the naive non-accumulative implementation has quadratic complexity in a functional language).
It is well known in the program inversion literature~\cite{DBLP:conf/pepm/MatsudaIN12,DBLP:conf/rta/NishidaV11}
that tail recursive functions (which are often needed for accumulation) are challenging to handle due to overlapping branch bodies.
The $\var{reverse}$ definition we use in the benchmark includes a small fix: it takes an additional parameter that represents the length of the list in the accumulation parameter. It is sufficient to guarantee the success of \SysName.




\paragraph{Class  \SomeLost} As we can see, \SysName~also performs well for this class. But as explained above, the success is conditional on the input/output examples that the $\PUT$ is required to satisfy. Take $\var{append}$ as an example, if the following example is included, which demands the second list being filled before the original first list is fully reconstructed, the synthesis will fail, as a solution must have a different recursion structure from that of the sketch.
\begin{center}
  \begin{tabular}{l|l|l|l}
    Original Source    & (Original View) & Updated View & Updated Source \\ \hline
    $([1,2,3,4], [5])$ & $[1,2,3,4,5]$   & $[6,2]$      & $([6], [2])$
  \end{tabular}
\end{center}
An interesting case is $\var{lines}$, which splits a string by $\texttt{'\textbackslash n'}$ to produce a list of strings. The synthesis becomes a lot harder when the examples (as seen in Table~\ref{table:examples-experiments}) require the preservation of the existence of the newline in the last position. This combined with structural
changes to the view list cannot be captured by the recursion structure of the
sketch, which explains the failure.

\paragraph{Class \AllLost}
Functions such as $\var{lookup}$ completely lose the source structures.
Consequently, {\SysName} will not be able to handle any example of structural changes. In the case of $\var{lookup}$, a structural change means that the view value is changed to another value associated to a different key in the source (as seen in Table~\ref{table:examples-experiments}). Just for demonstration, if only non-structural changes are considered, as in the following example
where the changed view does not switch to a different key,
{\SysName} will be able to successfully generate a program.
\begin{center}
  \begin{tabular}{l|l|l|l}
    Original Source                    & (Original View) & Updated View & Updated Source                    \\ \hline
    $([(1,10), (2,200), (3,33)],\, 2)$ & $200$           & $10$         & $([(1,10), (2,10), (3,33)],\, 2)$
  \end{tabular}
\end{center}
However, this is not interesting as the strength of HOBiT lies in its ability to
handle structural changes through branch switching.

\subsection{Larger and More Involved Example}\label{sec:experiment-complex}

Next, we evaluate {\SysName} on some larger examples, which are closer to realistic use cases. 
In particular, we look at two types of transformations: XML queries and string parsing.


\subsubsection{XML Transformations}\label{sec:XML_Transfromation}

We examined six queries from  XML Query Use Cases\footnote{\url{https://www.w3.org/TR/xquery-use-cases}} (``TREE'' Use Case).
Table~\ref{fig:detail-problem-xml} provides brief explanations for these queries and 
Figure~\ref{fig:xml-sample} shows the skeleton of the XML document used as the original source for them.
Such XML documents are represented in HOBiT by a rose-tree datatype. 
We ignored Document Type Definitions for simplicity---we could handle such constraints by fusing a partial identity function checking them to a $\GET$ function.
We also provided the constant ``title'' as an auxiliary component to our synthesis engine.

Table~\ref{fig:result-experiments-xml} contains the results of this experiment.
Column ``Updates'' indicates the updates of the source query triggered by the given input/output examples,
whereas columns ``LOC$_\mathrm{in}$'' and ``LOC$_\mathrm{syn}$'' denote the number of lines of code in the original and the synthesized query, respectively.
The results show that {\SysName} can synthesize fairly large
HOBiT programs. \revised{In particular, the number of AST nodes synthesised ranges from 73 (for Q4) to 471 (for Q5),
  corresponding to 17 lines of code for Q4 and 80 for Q5.
  \ifFullVersion
    (The programs are too large to be displayed in the main body of this paper.
    See Appendix~\ref{sec:concrete-input-output-for-q1} for the concrete input and output of {\SysName} for Q1,
    which serves as a representative of the six to illustrate their complexity.)
  \fi
}
The reason Q6 takes significantly more time than the rest
is that it assumes that each section has a title
element. Consequently, when handling insertion of sections, the
generated reconciliation function needs to construct a section with a
title. 


\begin{figure}
  \footnotesize
  \begin{alltt}
    <book><title>Data on the Web</title>
    <author>Serge Abiteboul</author><author>Peter Buneman</author><author>Dan Suciu</author>
    <section id="intro" difficulty="easy">
    <title>Introduction</title><p>\dots</p>
    <section><title>Audience</title><p>\dots</p></section>
    <section>
    <title>Web Data and the Two Cultures</title>
    <p>\dots</p>
    <figure height="400" width="400">
    <title>Traditional client/server architecture</title><image source="csarch.gif"/>
    </figure>
    <p>\dots</p>
    </section>
    </section>
    <section id="syntax" difficulty="medium">\dots</section>
    </book>
  \end{alltt}
  \caption{An XML document used as an original source for Queries Q1 to Q6.}
  \label{fig:xml-sample}
\end{figure}

\begin{table}[t]
  \caption{Explanations of the examined XML queries: the descriptions are quoted from XML Query Use Case, where ``Book1'' refers the source XML.}
  \label{fig:detail-problem-xml}\small
  \begin{tabular}{@{}c|p{12cm}@{}}
    Problem & Description (quoted)                                                                                                                                                                                                                                                \\\hline
    Q1      & ``Prepare a (nested) table of contents for Book1, listing all the sections and their titles. Preserve the original attributes of each <section> element, if any.''                                                                                                  \\[3pt]
    Q2      & ``Prepare a (flat) figure list for Book1, listing all the figures and their titles. Preserve the original attributes of each <figure> element, if any.''                                                                                                            \\[3pt]
    Q3      & ``How many sections are in Book1, and how many figures?''                                                                                                                                                                                                           \\[3pt]
    Q4      & ``How many top-level sections are in Book1?''                                                                                                                                                                                                                       \\[3pt]
    Q5      & ``Make a flat list of the section elements in Book1. In place of its original attributes, each section element should have two attributes, containing the title of the section and the number of figures immediately contained in the section.''                    \\[3pt]
    Q6      & ``Make a nested list of the section elements in Book1, preserving their original attributes and hierarchy. Inside each section element, include the title of the section and an element that includes the number of figures immediately contained in the section.''
  \end{tabular}
\end{table}

\begin{table}[t]
  \caption{The results of experiments for XML examples.}
  \label{fig:result-experiments-xml}\small
  \begin{tabular}{@{}c|rrlrrr@{}}
    Problem             & \(\mathrm{LOC}_\mathrm{in}\)  & \revised{\(\mathrm{AST}_\mathrm{in}\)} & Updates in I/O examples                    & Time (s)                         & \(\mathrm{LOC}_\mathrm{syn}\) & \revised{\(\mathrm{AST}_\mathrm{syn}\)} \\\hline
    \multirow{3}{*}{Q1} & \multirow{3}{*}{11}           & \multirow{3}{*}{\revised{136}}         & Add attribute(s)                           & \multirow{3}{*}{0.35\phantom{}}  & \multirow{3}{*}{42}           & \multirow{3}{*}{\revised{319}}          \\
                        &                               &                                        & Remove section(s)                          &                                  &                               &                                         \\
                        &                               &                                        & Change title(s) and attribute value(s)     &                                  &                               &                                         \\\hline
    \multirow{4}{*}{Q2} & \multirow{4}{*}{23}           & \multirow{4}{*}{\revised{199}}         & Remove figure(s)                           & \multirow{4}{*}{0.97\phantom{}}  & \multirow{4}{*}{69}           & \multirow{4}{*}{\revised{406}}          \\
                        &                               &                                        & Change title(s)                            &                                  &                               &                                         \\
                        &                               &                                        & Change attribute value(s)                  &                                  &                               &                                         \\
                        &                               &                                        & Add Attribute                              &                                  &                               &                                         \\\hline
    \multirow{3}{*}{Q3} & \multirow{3}{*}{20}           & \multirow{3}{*}{\revised{191}}         & Decrease figure count                      & \multirow{3}{*}{0.43\phantom{}}  & \multirow{3}{*}{63}           & \multirow{3}{*}{\revised{339}}          \\
                        &                               &                                        & Increase section count                     &                                  &                               &                                         \\
                        &                               &                                        & Decrease section count                     &                                  &                               &                                         \\\hline
    \multirow{2}{*}{Q4} & \multirow{2}{*}{9}            & \multirow{2}{*}{\revised{97}}          & Increase section count                     & \multirow{2}{*}{0.14}            & \multirow{2}{*}{26}           & \multirow{2}{*}{\revised{170}}          \\
                        &                               &                                        & Decrease section count                     &                                  &                               &                                         \\\hline
    \multirow{4}{*}{Q5} & \multirow{4}{*}{35}           & \multirow{4}{*}{\revised{342}}         & Decrease figure count                      & \multirow{4}{*}{1.2\phantom{0}}  & \multirow{4}{*}{115}          & \multirow{4}{*}{\revised{813}}          \\
                        &                               &                                        & Increase figure count                      &                                  &                               &                                         \\
                        &                               &                                        & Remove section(s)                          &                                  &                               &                                         \\
                        &                               &                                        & Change title(s)                            &                                  &                               &                                         \\\hline
    \multirow{4}{*}{Q6} & \multirow{4}{*}{\revised{31}} & \multirow{4}{*}{\revised{236}}         & Increase figure count(s)                   & \multirow{4}{*}{10\phantom{.00}} & \multirow{4}{*}{90}           & \multirow{4}{*}{\revised{530}}          \\
                        &                               &                                        & Decrease figure count(s)                   &                                  &                               &                                         \\
                        &                               &                                        & Add section(s) with title and figure count &                                  &                               &                                         \\
                        &                               &                                        & Change title(s)                            &                                  &                               &                                         \\\hline
  \end{tabular}
\end{table}

\subsubsection{Lexer and Parser}

We also examined a simple recursive decent (specifically, LL(1)) lexer and parser.
The lexer takes in strings (i.e., $\{ \texttt{(}, \texttt{)}, \texttt{S}, \texttt{Z}, \texttt{+} \}^{*}$)
and returns a sequence of tokens represented by the following datatype.
\begin{codemath}\bb
  \DATA~
  \con{Token} = \con{TNum} \A \con{Nat} \mid \con{LPar} \mid \con{RPar} \mid \con{Plus}
  \ee&&\end{codemath}
Note that natural numbers such as \texttt{S(S(Z))} are processed in this step.
Then, the parser takes in the output of the lexer, \ie, a sequence of the tokens above,
and returns an abstract syntax tree, according to the following grammar.
\[
  s ::= n \mid \texttt{(}s\texttt{)} \texttt{+} \texttt{(} s \texttt{)}
\]

The lexer and parser considered here are injective, which is uncommon in practice.
Typically, a lexer loses information about white spaces (layouting) and comments,
and a parser may remove syntactic sugars and redundant parentheses\footnote{A notable exception is the parser for GHC/Haskell, which keeps syntactic sugars and parentheses for better error messaging.}.
Sometimes, though, such lost information is attached to the abstract syntax trees,
making the parsing process injective~\cite{DBLP:conf/sle/JongeV11,DBLP:conf/scam/KortL03,DBLP:conf/pldi/PombrioK14}.

\begin{table}[t]
  \caption{Experimental results for the lexer and parser.}
  \label{table:results-lexer/parser}\small
  \begin{tabular}{l|rrlrrr}
           & \(\mathrm{LOC}_\mathrm{in}\) & \revised{\(\mathrm{AST}_\mathrm{in}\)} & \begin{tabular}{l}
      Updates
    \end{tabular} & Time (s) & \(\mathrm{LOC}_\mathrm{syn}\) & \revised{\(\mathrm{AST}_\mathrm{syn}\)} \\\hline
    Lexer  & \revised{15}                 & \revised{88}                           &
    \begin{tabular}{l}
      Change on natural numbers \\   remove/insert tokens
    \end{tabular}
           & 0.094                        & \revised{69}                           & \revised{265}
    \\
    Parser & \revised{10}                 & \revised{45}                           & \begin{tabular}{l}
      Replacement of whole AST and back
    \end{tabular}
           & 0.55\phantom{0}              & 29                                     & \revised{121}
  \end{tabular}
\end{table}

Table~\ref{table:results-lexer/parser} summarizes the experimental results.
%
In Figure~\ref{fig:synthesized-parser}, we provide the parser generated by {\SysName} as it is the more intricate of the two.
Notice that the $\con{LPAR}$ case in $\var{go}$ requires quite an involved reconciliation function.

\begin{figure}
  \setlength{\abovedisplayskip}{0pt}
  \setlength{\belowdisplayskip}{0pt}\footnotesize
  \setlength{\jot}{0pt}
  \begin{align*}
    \bb
    \var{pExp} :: {\BX{[\con{Token}]}} \to {\BX{\con{Exp}}}                                                                                                                                                                                                                                                                                                                                                                         \\
    \var{pExp} \A \var{ts} = \BLET~(e,[\,]) = \var{go} \A \var{ts}~\BIN~e                                                                                                                                                                                                                                                                                                                                                           \\[\blanklineskip]%
    \var{go} :: {\BX{[\con{Token}]}} \to \BX{(\con{Exp}, [\con{Token}])}                                                                                                                                                                                                                                                                                                                                                            \\
    \var{go} \A \var{ts} =
    \BCASE~\var{ts}~\BOF~                                                                                                                                                                                                                                                                                                                                                                                                           \\
    \ \ \
    \bbt
    \con{TNum} \A n:r & \to                                                                                                                                                                                                       & \bx{(}\bx{\con{ENum}} \A n\,\bx{,}\,r\bx{)}                                                                                                                                                     \\
                      &                                                                                                                                                                                                           & \bbt \BWITH                                                               & \lambda v.\, \CASE~v~\OF~\{ (\con{ENum}\A\dontcare,\dontcare)\to \con{True};\;\dontcare \to \con{False}\}           \\
    \BBY              & \lambda s.\lambda v.\, \CASE~v~\OF~\{ (\con{ENum}\A a,\dontcare)\to \con{TNum} \A a:s\} \ee                                                                                                                                                                                                                                                                                                                 \\
    \con{LPAR}:r_1    & \to                                                                                                                                                                                                       & \BLET~(e_1,\con{RPar}:\con{Plus}:\con{LPar}:r_2)=\var{go}\A\var{r_1}~\BIN                                                                                                                       \\
                      &                                                                                                                                                                                                           & \BLET~(e_2,\con{RPar}:\con{Plus}:\con{LPar}:r_3)=\var{go}\A\var{r_2}~\BIN                                                                                                                       \\
                      &                                                                                                                                                                                                           & \quad \bx{(}\bx{\con{EAdd}} \A e_1 \A e_2\,\bx{,}\, r_3\bx{)}                                                                                                                                   \\
                      &                                                                                                                                                                                                           & \bbt \BWITH                                                               & \lambda v.\,\CASE~v~\OF~\{ (\con{EAdd}\A\dontcare\A\dontcare,\dontcare)\to \con{True};\;\dontcare \to \con{False}\} \\
    \BBY              & \lambda s.\lambda v.\,\CASE~v~\OF~\{ (\con{EAdd} \A \dontcare \A \dontcare, \dontcare) \to \con{LPar} : \con{TNum}\A\con{Z} : \con{RPar}: \con{Plus} :\con{LPar} : \con{TNum}\A\con{Z}:\con{RPar}:s\} \ee
    \ee
    \ee\end{align*}
  \caption{Synthesized Bidirectional Parser}
  \label{fig:synthesized-parser}
\end{figure}

\subsection{Comparison with \textsc{Smyth}}\label{sec:experiment-comparative}
A fair comparison with other synthesis systems is not always easy
due to the different set-ups. For example, we
cannot compare directly with Optician~\cite{DBLP:journals/pacmpl/MiltnerFPWZ18,DBLP:journals/pacmpl/MainaMFPWZ18,DBLP:journals/pacmpl/MiltnerMFPWZ19}, the state of the art lens synthesizer, as its inputs and outputs
are too different from ours (see Section~\ref{sec:related-work} for a non-experimental comparison).


Instead, we pick \textsc{Smyth}~\cite{DBLP:journals/pacmpl/LubinCOC20}, a state-of-the-art synthesis tool that synthesizes unidirectional
programs from sketches and input/output
examples---a set-up that is similar to ours.
We provide to \textsc{Smyth} hand-written sketches of $\PUT$ in the form of ``base case sketches''~\cite{DBLP:journals/pacmpl/LubinCOC20}, which are incomplete programs for which the step case branches are left as holes
while the base case branches are pre-filled, and the same input/output examples as the experiments in Table~\ref{table:examples-experiments}.
We omit the round-tripping requirement for \textsc{Smyth} and only
check whether the tool is able to produce $\PUT$ functions that satisfy the
input/output
examples.

Table~\ref{table:result-comparative} shows the results of the comparison (with more details in the artifact \footnote{\url{https://doi.org/10.5281/zenodo.5494504}} or the repository\footnote{\url{https://github.com/masaomi-yamaguchi/synbit}}).   
  {\SysName} successfully synthesized 13 out of 15 cases, whereas \textsc{Smyth} succeeded only in 7 cases.
We believe that the main reason for the difference is the  required $\PUT$ functions tend to be quite complex, usually more so than their corresponding $\GET$.
It is worth noting that \textsc{Smyth} succeeded for $\var{lookup}$, where {\SysName} failed.
For this particular case, a $\PUT$ program that conforms to the input/output example is represented by the key-value-flipped version of $\GET$, which was ruled out in {\SysName} by a sketch.



\begin{table}[t]
  \caption{Results of comparative experiments with \textsc{Smyth}: ``No'' means that \textsc{Smyth} reported failure in 10 min.}
  \label{table:result-comparative}\small
  \begin{tabular}[t]{@{}lc|cccc@{}}
    Problem                  & Class         & {\SysName} & \textsc{Smyth} \\\hline
    $\var{double}$           & \AllPreserved & Yes        & Yes            \\
    $\var{uncurryReplicate}$ & \AllPreserved & Yes        & Yes            \\
    $\var{mapNot}$           & \AllPreserved & Yes        & Yes            \\
    $\var{mapReplicate}$     & \AllPreserved & Yes        & Yes            \\
    $\var{snoc}$             & \AllPreserved & \bf Yes    & No             \\
    $\var{length}$           & \AllPreserved & Yes        & Yes            \\
    $\var{lengthTail}$       & \AllPreserved & Yes        & Yes            \\
    $\var{reverse}$          & \AllPreserved & \bf Yes    & No             \\
  \end{tabular}\hspace{10pt}
  \begin{tabular}[t]{@{}lc|cccc@{}}
    Problem           & Class         & {\SysName} & \textsc{Smyth} \\\hline
    $\var{mapFst}$    & \AllPreserved & \bf Yes    & No             \\
    $\var{add}$       & \SomeLost     & \bf Yes    & No             \\
    $\var{append}$    & \SomeLost     & \bf Yes    & No             \\
    $\var{appendBc}$  & \SomeLost     & \bf Yes    & No             \\
    $\var{professor}$ & \SomeLost     & \bf Yes    & No             \\
    $\var{lines}$     & \SomeLost     & Timeout    & Timeout        \\
    $\var{lookup}$    & \AllLost      & Timeout    & \bf Yes        \\
  \end{tabular}
\end{table}



\section{Related work}
\label{sec:related-work}



\newcommand{\concatdot}{\mathbin{\texttt{.}}}


\paragraph{Optician}
Optician~\cite{DBLP:journals/pacmpl/MiltnerFPWZ18,DBLP:journals/pacmpl/MainaMFPWZ18,DBLP:journals/pacmpl/MiltnerMFPWZ19} is the state-of-the-art framework for synthesizing lenses~\cite{FGMPS07,HofmannPW11,DBLP:conf/popl/BohannonFPPS08,DBLP:conf/icfp/FosterPP08}.
Both their framework and ours implicitly guarantee the round-tripping properties
by using bidirectional programming languages (lenses/HOBiT) as targets. However, a direct comparison of performance is difficult due
to the very different set-ups. Their target lenses are specialized for string transformations, while HOBiT
considers general datatypes. And correspondingly, the core of their input specification is
regular expressions describing
data formats, while that of ours is standard functional programs serving as sketches. 
\begin{revisedblock}
Due to such differences in set-ups, even though we could translate a specification for Optician (regular expressions and input/output examples) to one for {\SysName} (a $\GET$ program and input/output examples), such a translation would  
involve many arbitrary choices (especially the choice of a  $\GET$ for Synbit) that affect synthesis, 
effectively ruling out a meaningful comparison (see Appendix~\ref{sec:why-not-have-side-by-side-comparison-with-optician}
for an illustrative example on the difficulty). 
\end{revisedblock}

Despite the very different approaches,
it is interesting to observe a common design principle shared by both: leveraging the strengths of the underlying
bidirectional languages. Optician's regular-expression-based specification matches perfectly with the simplicity of the lens languages and their
close connection to advanced types, while {\SysName} takes full advantage of HOBiT's alignment to conventional functional programming.
On a more technical note, Optician~\cite{DBLP:journals/pacmpl/MiltnerMFPWZ19} is able to prioritize
the generated programs by quantitative information flow. It is not clear how this may be used in {\SysName} as the
computation of the quantitative information flow will be difficult for a language with arbitrary recursion.

\paragraph{Other synthesis efforts for bidirectional programming}

In a vision paper, \citet{DBLP:conf/pepm/Voigtlander12} suggests some directions
of synthesizing bidirectional programs from $\GET$ programs by leveraging the round-tripping properties.
Specifically, he suggests using the round-tripping properties to generate input/output examples for synthesis.
Using \ref{law:acceptability}, if one can generate $s$ in some ways (assuming the totality of $\GET$), then
examples of backward behavior may arise from $\PUT \A (s, \GET \A s) = s$. But a naive application of this
without considering \ref{law:consistency}
may result in incorrect $\PUT$ behavior such as $\PUT \A (s,v) = s$. To remedy the situation, Voigtl\"ander suggests
restricting $\PUT$ to use the second argument $v$; \ie, the argument must be relevant in the sense of relevant typing.
Voigtl\"ander also suggests using \ref{law:consistency} to restrict the form of $\PUT$ to satisfy $\PUT \A (s,v) \in \GET^{-1}(v)$,
which is indeed effective for simple $\GET$s such as $\GET = \var{head}$ (in this case the right-hand side must have the form of $v : \dontcare$).
However, synthesis in this direction does not guarantee correctness with respect to round-tripping, and
an additional verification process will then be needed.

The PINS framework~\cite{DBLP:conf/pldi/SrivastavaGCF11} applies path-based synthesis to program inversion, a program
transformation that derives the inverse of an (injective) program. 
The path-based synthesis was able to derive inverses for involved programs such as LZ77 and LZW compression which the other existing inversion methods at the present time cannot handle.
However, since it focuses only on a finite number of paths, the system does not guarantee correctness: the resulting programs may not always be inverses.
PINS also uses sketches and component functions given by users.

\begin{revisedblock}
\paragraph{Program inversion}
Program inversion is a technique related to bidirectional programming, but there are important differences. 
In program inversion, input programs are expected to be injective and thus serve as complete specifications,
which is not the case in bidirectional programming. As a result, in {\SysName} input/output examples are used 
to further specify the required backward behavior. 
Despite the differences, program inversion and bidirectional programming do share some common techniques.
For example, using postconditions (as exit conditions in HOBiT) to determine control flows (especially branches) in inverses is a very common approach in the literature~\cite{Korf81,Janus,YokoyamaAG08,YoAG11,Gries81,GlKa05,MMHT10}.
A more interesting connection is the concept of \emph{partial} inversion~\cite{NishidaSS05}, which uses 
binding-time analysis before inversion so that the inverses can use static data as inputs as well. 
Types in HOBiT can be seen as binding time where non-$\BX{}$-types are seen as static, 
and our type-directed sketch generation (Section~\ref{sec:synthesis-template-generation}), with the lazy nondeterministic generation,
can be viewed as a type-based binding-time analysis~\cite{GoJo91}.
The idea of partial inversion is further extended so that the return values of inverses are treated as ``static inputs'' as well~\cite{Almendros-JimenezV06}, and
the \emph{pin operator}~\cite{DBLP:journals/pacmpl/MatsudaW20} is proposed  to capture such a behavior 
in an invertible language. 
However, the utility of the operator in bidirectional programming rather than invertible programming is still under exploration, 
and thus our current synthesis method does not include it. 
\end{revisedblock}

\paragraph{Bidirectionalization}

Bidirectionalization is a program transformation that derives a bidirectional
transformation from a unidirectional transformation.
In a sense, this can be seen as a simple type of synthesis.  
\citet{MatsudaHNHT07}, based on the constant-complement view updating~\cite{BaSp81},
analyze injectivity (information-loss) of a program and then derive a complement by gathering lost information to obtain a bidirectional version.
This method requires a strong restriction on input programs for effective analysis: they must be affine (no variables can be used more than once) and treeless~\cite{DBLP:journals/tcs/Wadler90} (only variables can be arguments of functions) so that the injectivity analysis becomes exact.
\citet{DBLP:conf/popl/Voigtlander09} makes use of parametricity~\cite{DBLP:conf/fpca/Wadler89,DBLP:conf/ifip/Reynolds83} to interpret polymorphic functions as bidirectional transformations. The technique is restricted to polymorphic functions. Probably more importantly, it can only handle non-structural updates---the equivalent of HOBiT without the ability of branch switching. Several extensions of the idea have been proposed~\cite{DBLP:journals/jfp/VoigtlanderHMW13, DBLP:conf/icfp/MatsudaW15,DBLP:journals/jfp/Matsuda018}. In general, bidirectionalization is
far less expressive than the state-of-the-art synthesis frameworks such as Optician/lenses and \SysName/HOBiT.

\paragraph{General program synthesis}

A popular direction in program synthesis that inspired our work is
program sketching, where programmers express their insights about a
program by writing sketches, i.e., partial programs encoding the
structure of a solution while leaving its low-level details
unspecified in the form of holes
\cite{DBLP:conf/aplas/Solar-Lezama09}. As opposed to our technique,
\citet{DBLP:conf/aplas/Solar-Lezama09} can only be applied to integer
benchmarks and does not support other data types such as lists or
trees. Also, it mostly focuses on properties, rather than examples, by
relying on Counterexample Guided Inductive Synthesis
(CEGIS)~\cite{DBLP:conf/pldi/Solar-LezamaJB08}, where a candidate
solution is iteratively refined based on counterexamples provided by a
verification technique. There is actually a large body of works based
on the CEGIS architecture
\cite{DBLP:journals/toplas/DavidKKL18,DBLP:conf/esop/DavidKL15,DBLP:conf/cav/AbateBCCDKKP17,DBLP:conf/icse/JhaGST10,DBLP:conf/oopsla/KneussKKS13,DBLP:conf/cav/AbateDKKP18}. Often, such
approaches expect formal specifications describing the behavior of
the target program, which 
can be difficult to write or expensive to check against using automated verification techniques.
Conversely, our specification consists of the unidirectional program and input/output
examples, which we believe is intuitive and easy to use, without
requiring prior understanding of logic.

The original work on program sketching has inspired a multitude of
follow-up directions.  Some of the most related to our work are
\citet{DBLP:conf/sfp/Katayama05,DBLP:conf/pldi/FeserCD15,DBLP:conf/pldi/OseraZ15,DBLP:journals/pacmpl/LubinCOC20},
which, similarly to our technique, are type-directed and guided by input/output examples. As
opposed to these approaches, we exploit information about the
unidirectional program in order to prune the search space for the
bidirectional correspondent. As shown in our experimental
evaluation, simply applying synthesis techniques designed for
unidirectional code is not effective. 

Another direction that inspired us is that of component-based
synthesis~\cite{DBLP:conf/icse/JhaGST10,DBLP:conf/popl/FengM0DR17}, where the target program is generated by composing components from a library.  Similarly to these approaches, we
use a given library of components as the building blocks of our
program generation approach.

\begin{revisedblock}
\paragraph{Equivalence reduction}
Program synthesis techniques make use of equivalence reduction in order to
reduce the number of equivalent programs that get explored.
For example, \citet{DBLP:conf/cav/AlbarghouthiGK13} 
prune the search space using observational equivalence with respect to a set of input/output examples, i.e., two programs are considered to be in the same equivalence class if, for all given inputs in the set of input/output examples, they produce the same outputs. Alternatively, \citet{DBLP:conf/vmcai/SmithA19} generate only programs in a specific normal form, where term rewriting is used to transform a program into its normal form. In~\cite{DBLP:journals/corr/KoukoutosKK16}, Koukoutos et al. make use of attribute grammars to only produce certain types of expressions in their normal form, thus skipping other expressions
that are syntactically different, yet semantically equivalent.
In our work, we found that the lightweight heuristics described in Section~\ref{sec:heuristics}
worked well. However, we do plan on exploring some of the equivalence
reduction techniques discussed here as future work.
\end{revisedblock}

\section{Conclusion}
\label{sec:conclusion}

We proposed a synthesis method for bidirectional transformations, whose novelty lies in the use of $\GET$ programs as sketches.
We described the idea in detail and implemented it in a prototype system {\SysName}, where
lazy nondeterministic generation has played an important role.
Through the experiments, we demonstrated the effectiveness of the proposed method and clarified its limitations.

A future direction is to make use of program analysis and verification techniques in the synthesis of exit conditions.
This would enable us to guarantee stronger soundness as discussed in Section~\ref{sec:soundness-incompleteness}.
Another future direction is to extend the target language (HOBiT) based on our experience in order to synthesize more bidirectional
transformations. 

\begin{acks}

We thank Eijiro Sumii and Oleg Kiselyov for their helpful and instructive comments on an earlier stage of this research,
and Hiroshi Unno for fruitful discussions on future directions.
This work was partially supported by \grantsponsor{JSPS}{JSPS}{https://www.jsps.go.jp/} KAKENHI Grant Numbers \grantnum{JSPS}{15H02681}, \grantnum{JSPS}{19K11892} and \grantnum{JSPS}{20H04161},
JSPS Bilateral Program, Grant Number \grantnum{JSPS}{JPJSBP120199913}, \grantsponsor{Kayamori}{the Kayamori Foundation of Informational Science Advancement}{http://www.kayamorif.or.jp/}, \grantsponsor{EPSRC}{EPSRC}{https://epsrc.ukri.org/} Grant \emph{EXHIBIT: Expressive High-Level Languages for Bidirectional Transformations} (\grantnum{EPSRC}{EP/T008911/1}), \grantsponsor{RoyalSociety}{Royal Society}{} Grant \emph{Bidirectional Compiler for Software Evolution} (\grantnum{RoyalSociety}{IES\textbackslash R3\textbackslash 170104}), and Royal Society University Research Fellowship \emph{On Advancing Inductive Program Synthesis} (\grantnum{RoyalSociety}{UF160079}).

\end{acks}

\ifFullVersion
  \bibliographystyle{ACM-Reference-Format}
  \bibliography{main}
\appendix

\section{Appendix}

\begin{revisedblock}
\subsection{More Discussion on the Difficulty of Side-by-Side Comparison with Optician}
\label{sec:why-not-have-side-by-side-comparison-with-optician}

Due to the very different set-ups, a side-by-side comparison of \SysName{} and Optician
is problematic. The 
arbitrary choices required to bridge the gap make a fair comparison out of reach. 
We illustrate this problem with an example (\texttt{extr-fname.boom} in Fig.~\ref{fig:extr-fname}) taken from the artifact associated with the Optician papers~\cite{DBLP:journals/pacmpl/MiltnerMFPWZ19,DBLP:journals/pacmpl/MiltnerFPWZ18}.

The specification describes the task of separating a path into a file and a directory path. 
As one can see, most of the code is devoted to specifying the input and output formats (\var{NONEMPTYDIRECTORY} and \var{FILEANDFOLDER}). 
The input/output examples are specified by the \key{using} clause: $\key{createrex}$ provides an example of how a source is related to a view.
Note that this specification targets the synthesis of bijective transformations; so the backward behavior does not require the original source. 


Let us consider how we can encode this specification to be used by \SysName.
%
As a first step, we need to decide the types for inputs and outputs. One candidate is using strings (lists of characters in HOBiT). In such a case, it is natural to divide the task into three subtasks: (1) parsing (of type $\con{String} \to S$), (2) core transformation (of type $S \to T$), and (3) printing (of type $T \to \con{String}$), such that the interesting computation is done in the middle. 
For the comparison to Optician, it makes sense to only consider the core transformation; parsing and printing are coupled with lens combinators used in Optician and are not synthesized separately from the core transformation.

We then need to decide the domain ($S$) and range ($T$) of the core transformation.
One option is to use $S = (\con{NonEmpty} \A \con{String}, \con{Bool})$ and $T = (\con{String}, \con{Bool}, [\con{String}])$, where:
\begin{codemath}\bb
\TYPE~\con{NonEmpty} \A a = (a, [a]) \qquad \LCOMMENT{head-biased non-empty lists}
\ee\end{codemath} 
Another option is to use datatypes that mirror the structure of regular expressions, such as: 
\begin{codemath}\bb
\DATA~\con{LC} = \con{LA} \mid \con{LB} \mid \dots \mid \con{LZ} \\
\DATA~\con{UC} = \con{UA} \mid \con{UB} \mid \dots \mid \con{UZ} \\
\DATA~\con{C} = \con{Lower} \A \con{LC} \mid \con{Upper} \A \con{UC} \mid \con{UnderScore} \mid \con{Dot} \mid \con{Hyphen} \\
\TYPE~\con{LocalFolder} = (\con{C}, [\con{C}]) \\
\TYPE~\con{Directory} = (\con{Bool}, [\con{LocalFolder}]) \\
\TYPE~\con{NonEmptyDirectory} = (\con{Bool}, \con{LocalFolder}, [\con{LocalFolder}]) \\
\TYPE~\con{FileAndFolder} = (\con{LocalFolder}, \con{Directory})
\ee\end{codemath}
In this particular case, the choice between the two does not affect the core transformation part much; in both cases, it essentially performs a
transformation from head-biased nonempty lists to last-biased ones, with some arrangement of products.  
So, one can think that the essential part of this transformation is a function of type $\var{headBiased2LastBiased} :: (A,[A]) \to ([A],A)$ for some concrete type $A$. Note that abstracting the concrete type $A$ by a type variable $a$ here gives us the information that the components of the lists are not touched by the transformation. 

\begin{figure}
\begin{revisedblock}%
\footnotesize\begin{codemath}\bb
\LET~\var{LOWERCASE} : \key{regexp} = \texttt{"a"} \mid \texttt{"b"} \mid \dots \texttt{(* omitted *)} \dots  \mid \texttt{"z"}\\
\LET~\var{UPPERCASE} : \key{regexp} = \texttt{"A"} \mid \texttt{"B"} \mid \dots \texttt{(* omitted *)} \dots  \mid \texttt{"Z"}\\[\blanklineskip]%
\LET~\var{LOCALFOLDER} : \key{regexp} = \\
\qquad \bbt (\var{LOWERCASE} \mid \var{UPPERCASE} \mid \texttt{"\_"} \mid \texttt{"."} \mid \texttt{"-"})\\
  {}\concatdot (\var{LOWERCASE} \mid \var{UPPERCASE} \mid \texttt{"\_"} \mid \texttt{"."} \mid \texttt{"-"}){*}\ee\\
\LET~\var{DIRECTORY} : \key{regexp} = 
  (\texttt{"/"} \mid \texttt{""}) \concatdot (\var{LOCALFOLDER} \concatdot \texttt{"/"}){*}\\
\LET~\var{NONEMPTYDIRECTORY} : \key{regexp} = {}\\
\qquad  (\texttt{"/"} \mid \texttt{""}) \concatdot \var{LOCALFOLDER} \concatdot (\texttt{"/"} \concatdot \var{LOCALFOLDER}){*}\\
\LET~\var{FILEANDFOLDER} : \var{regexp} = {}\\
\qquad  \texttt{"file: "} \concatdot \var{LOCALFOLDER} \concatdot \texttt{"\textbackslash{}nfolder: "} \concatdot \var{DIRECTORY}\\
\LET~\var{extract\_file} : (\key{lens}~\key{in}~\var{NONEMPTYDIRECTORY} \Leftrightarrow \var{FILEANDFOLDER}) = {}\\
\quad 
 \bbt
  \key{synth}~\var{NONEMPTYDIRECTORY} \Leftrightarrow \var{FILEANDFOLDER}\\
  \key{using}~\{\\
  \quad \bbt
       \key{createrex} (\bbt\texttt{"/Users/amiltner/lens/tests/flashfill/extract-filename.txt"}, \\
                            \texttt{"file: extract-filename.txt\textbackslash{}nfolder: /Users/amiltner/lens/tests/flashfill/"}),\ee \\
       \key{createrex} (\bbt\texttt{"tests/flashfill/extract-filename.txt"},\\
                            \texttt{"file: extract-filename.txt\textbackslash{}nfolder: tests/flashfill/"})\ee\\
       \ee\\
  \} 
 \ee
\ee\end{codemath}%
\end{revisedblock}%
\caption{\revised{\texttt{extr-fname.boom} for bijective-lens synthesis (excerpt) }}
\label{fig:extr-fname}
\end{figure}


The above set-up may sound reasonable but actually omits important internal details.
Optician internally tries to expand 
$r*$ into either $rr{*} | \varepsilon$ or $r{*}r|\varepsilon$ nondeterministically~\cite{DBLP:journals/pacmpl/MiltnerMFPWZ19}, which 
eventually transforms $r r{*}$ (head-biased non-empty lists) into $r (r{*}r | \varepsilon) = rr{*}r | r$ (one-step expansions of last-biased nonempty lists). 
The core transformation involves no structural transformations after this expansion.
However, this expansion of the Kleene star conflicts with {\SysName}, where the input and output types have to be fixed beforehand. 
Optician dynamically searches for a suitable-for-synthesis regular expression among equivalent ones mainly by converting them to ``sum-of-product'' forms and then by applying the expansion above~\cite{DBLP:journals/pacmpl/MiltnerMFPWZ19}. 


Trying to give a concrete definition of the transformation is even more problematic,
with semantically equivalent definitions having very different effects on synthesis. 
For example, if we define $\var{headBiased2LastBiased} :: (A,[A]) \to ([A],A)$ as the following:
\begin{codemath}\bb
\var{headBiased2LastBiased} \A (a,\var{as}) = \var{initlast} \A a \A \var{as}\\
\bbt
\var{initlast} \A a \A [\,] &= ([\,], a) \\ 
\var{initlast} \A a \A (b : \var{bs}) &= 
 \LET~(i,l) = \var{initlast} \A b \A \var{bs}~\IN~(a : i, l ) 
\ee
\ee\end{codemath}
{\SysName} has no problem in synthesizing a bidirectional version of it. 
On the other hand, the following equivalent definition does not work well. 
\begin{codemath}\bb
\var{headBiased2LastBiased} \A (a,\var{as}) = (\var{init} \A a \A \var{as} , \var{last} \A a \A \var{as}) \\
\bbt 
\var{init} \A a \A [\,] &= [\,] \\
\var{init} \A a \A (b : \var{bs}) &= a : \var{init} \A b \A \var{bs} 
\ee\\
\bbt 
\var{last} \A a \A [\,] &= a \\
\var{last} \A a \A (b : \var{bs}) &= \var{last} \A b \A \var{bs} 
\ee
\ee\end{codemath}
The reason for this is that the bijective transformation is separated into non-injective components $\var{init}$ and $\var{last}$.
Non-injectivity is usually not a problem as {\SysName} is designed to handle them with $put$. But in this case, the information that the 
non-injective functions are combined to form a bijection is lost in the separation, which restricts the updates that the backward 
function may handle. {\SysName} will (correctly) insist that the input data discarded by $\var{init}$\slash$\var{last}$ cannot be 
changed in the backward execution (otherwise, the round-tripping properties will be (locally) violated), which in this case results in a useless bidirectional program
that rejects all changes (and of course the synthesis fails at this point as the input/output examples cannot be satisfied).

In a similar manner, the opposite direction of encoding {\SysName} examples in Optician is also problematic. 
A lot of cases will simply fail to translate, and for the rest, particular ways of encoding are required for Optician to work well. 
Due to this, a side-by-side comparison of the two systems will be forced and unlikely to produce meaningful results. 


It is apparent that Optician and {\SysName} occupy very different parts of the synthesis design space. 
This difference is driven by the differences in the underlying languages they target: lenses vs.\ HOBiT. 
Lenses are tricky to program with, but the language itself is very simple; it, therefore, makes sense to 
have a separate specification system that is removed from the target implementation.  
In contrast, HOBiT focuses more on programmability, and the specification system may naturally 
take advantage of the fact. In a sense, lenses may be considered to benefit more from 
synthesis, as it relieves the need to program directly in them. 
On the other hand, {\SysName} demonstrates the impact of the language design: it not only improves programmability
but also enables effective synthesis methods.

\end{revisedblock}

\ifFullVersion
\subsection{Full Definition of Typing Relation}
\label{sec:hobit-typing-complete}

The complete definition of (our fragment of) HOBiT's typing relation is given in Fig.~\ref{fig:hobit-typing}.
The definition assumes constructors have simple types $\con{C} : A_1 \to \dots \to A_n \to A$, which
may be instantiation results of their (rank-1) polymorphic types (such as list constructors $(:)$ and $[\,]$).

\begin{figure}[tb]
  \setlength{\abovedisplayskip}{0pt}
  \setlength{\belowdisplayskip}{0pt}
  \setlength{\jot}{3pt}
  \begin{flushleft}\fbox{$\Gamma; \Delta \vdash e : A$}\end{flushleft}
  \begin{gather*}
    \infer
    { \Gamma; \Delta \vdash x : A  }
    { \Gamma(x) = A }
    \quad
    \infer
    { \Gamma; \Delta \vdash x : \BX{\tau}}
    { \Delta(x) = \tau }
    \quad
    \infer
    { \Gamma; \Delta \vdash \lambda x.e : A \to B }
    { \Gamma, x:A; \Delta \vdash e : B }
    \quad
    \infer
    { \Gamma; \Delta \vdash e_1 \A e_2 : B }
    { \Gamma; \Delta \vdash e_1 : A \to B
      \quad
      \Gamma; \Delta \vdash e_2 : A }
    \\
    \infer
    {\Gamma; \Delta \vdash \con{C} \A \V{e}: A }
    { \{ \Gamma; \Delta \vdash e : A_i \}_i \quad \con{C} : A_1 \to \dots \to A_n \to A   }
    \quad
    \infer
    { \Gamma; \Delta \vdash \CASE~e_0~\OF~\{ p_i \to e_i \}_i : B }
    { \Gamma; \Delta \vdash e_0  : A
      \quad
      \{ \Gamma_i \vdash p_i : A \quad \Gamma, \Gamma_i ; \Delta \vdash e_i : B \}_i
    }
    \\
    \infer
    { \Gamma ; \Delta \vdash \LIFTU{e} : \BX{\tau} }
    { \Gamma ; \Delta \vdash e : \tau }
    \quad
    \infer
    { \Gamma ; \Delta \vdash \bxcon{C} \A \V{e}: \BX{\tau} }
    { \{ \Gamma; \Delta \vdash e_i : \BX{\tau_i} \}_i
      \quad
      \con{C} : \tau_1 \to \dots \to \tau_n \to \tau
    }
    \\
    \infer
    { \Gamma; \Delta \vdash \BCASE~e_0~\BOF~\{ {{p_i}} \to e_i~\BWITH~e'_i~\BBY~e''_i \}_i : \BX{\sigma} }
    { \Gamma; \Delta \vdash e_0 : \BX{\tau}
      \quad
      \{ \Delta_i \vdash p_i : A \quad \Gamma ; \Delta, \Delta_i \vdash e_i : B \quad \Gamma; \Delta \vdash e'_i : \sigma \to \con{Bool} \quad \Gamma; \Delta \vdash e''_i : \tau \to \sigma \to \tau  \}_i
    }
  \end{gather*}
  \vspace*{0.3\baselineskip}
  \begin{flushleft}\fbox{$\Gamma \vdash p : A$}\end{flushleft}
  \begin{gather*}
    \infer
    { x: A \vdash x : A }
    {}
    \quad
    \infer
    {\Gamma_1,\dots,\Gamma_n \vdash \con{C} \A \V{p} : A  }
    { \{ \Gamma_i \vdash p_i : A_i \}
      \quad
      \con{C} : A_1 \to \dots \to A_n \to A
    }
  \end{gather*}
  \vspace*{0.3\baselineskip}
  \begin{flushleft}\fbox{$\Gamma \vdash P$}\end{flushleft}
  \begin{gather*}
    \infer
    { \Gamma \vdash f_1 = e_1; \dots ; f_n = e_n  }
    { \{ \Gamma ; \emptyset \vdash e_i : \Gamma(f_i) \}_{i} }
  \end{gather*}
  \caption{Typing rules: $\Delta \vdash p : \sigma$ is defined similarly to $\Gamma \vdash p : A$ but asserts that the resulting environment is actually an bidirectional one and every type that occurs in the derivation is a $\sigma$-type.}
  \label{fig:hobit-typing}
\end{figure}
\fi

\ifFullVersion
\subsection{Full Definition of Sketch Generation Rules}
\label{sec:template-generation-rules-full}

Figure~\ref{fig:gen-template} provides the sketch generation rules used in Section~\ref{sec:synthesis-template-generation}.
For simplicity of presentation, in Section~\ref{sec:synthesis} we did not explicitly capture the type of the code to be generated in the specialized holes as it can be recovered from the sketch and typing environment $\Gamma$. However, here we make it explicit by augmenting
both the exit condition hole and the reconciliation hole with
the typing environment $\Gamma$ and the type of the code to be
synthesized: $\EHole{\Gamma}{\sigma \to \con{Bool}}{e}$ and
$\RHole{\Gamma}{\sigma_0 \to \sigma \to \sigma_0}{p}{e}$.

\begin{figure}[t]
  \setlength{\abovedisplayskip}{0pt}
  \setlength{\belowdisplayskip}{0pt}
  \setlength{\jot}{10pt}
  \small
  \begin{gather*}
    \ninfer
    {G-!}
    { \GenTmp{\Gamma'}{\Delta'}{\BX{\tau}}{\Gamma}{e}{\tau}{\LIFTU{e'}} }
    { \GenTmp{\Gamma'}{\Delta'}{\tau}{\Gamma}{e}{\tau}{e'} }
    \quad
    \ninfer
    {G-UVar}
    { \GenTmp{\Gamma'}{\Delta'}{A'}{\Gamma}{x}{A}{x}}
    { x:A\in\Gamma\quad x:A'\in\Gamma' }
    \quad
    \ninfer
    {G-BVar}
    { \GenTmp{\Gamma'}{\Delta'}{\BX{\sigma}}{\Gamma}{x}{\sigma}{x} }
    { x:\sigma\in\Gamma\quad x:\sigma\in\Delta' }
    \\
    \ninfer
    {G-Abs}
    { \GenTmp{\Gamma'}{\Delta'}{(A_1'\to A_2')}{\Gamma}{\lambda x.e}{A_1\to A_2}{\lambda x.e'} }
    { \GenTmp{(\Gamma',x:A_1')}{\Delta'}{A_2'}{(\Gamma,x:A_1)}{e}{A_2}{e'} }
    \\
    \ninfer
    {G-App}
    { \GenTmp{\Gamma'}{\Delta'}{A'}{\Gamma}{e_1 \A e_2}{A}{e_1' \A e_2'} }
    { \begin{tarray}{c}
        \GenTmp{\Gamma'}{\Delta'}{(A_2'\to A')}{\Gamma}{e_1}{(A_2\to A)}{e_1'} \quad
        A_2 \leadsto B_2 \quad
        \GenTmp{\Gamma'}{\Delta'}{B_2}{\Gamma}{e_2}{A_2}{e_2'}
      \end{tarray}
    }
    \\
    \ninfer
    {C-Con}
    {
      \GenTmp{\Gamma'}{\Delta'}{A}{\Gamma}{\con{C} \A \V{e}}{A}{\con{C} \A \V{e'}}
    }
    {
      \con{C} : A_1 \to \dots \to A_n \to A \quad
      \{ \GenTmp{\Gamma'}{\Delta'}{A_i}{\Gamma}{e_i}{A_i}{e'_i}  \}_i
    }
    \\
    \ninfer
    {C-BCon}
    {
      \GenTmp{\Gamma'}{\Delta'}{\BX{\tau}}{\Gamma}{\con{C} \A \V{e}}{\tau}{\bxcon{C} \A \V{e'}}
    }
    {
      \con{C} : \tau_1 \to \dots \to \tau_n \to \tau \quad
      \{ \GenTmp{\Gamma'}{\Delta'}{\BX{\tau_i}}{\Gamma}{e_i}{\tau_i}{e'_i}  \}_i
    }
    \\
    \ninfer
    {G-Case}
    { \GenTmp{\Gamma'}{\Delta'}{A'}{\Gamma}{\CASE~e_0~\OF~\{ p_i \to e_i \}_i}{A}{\CASE~e_0'~\OF~\{ p_i \to e_i' \}_i} }
    { \begin{tarray}{c}
        \GenTmp{\Gamma'}{\Delta'}{A_0'}{\Gamma}{e_0}{A_0}{e_0'} \quad
        A_0 \leadsto A_0' \\
        \{ \Gamma_i\vdash p_i: A_0 \quad
        \Gamma'_i \vdash p_i : A_0' \quad
        \GenTmp{(\Gamma',\Gamma_i')}{\Delta'}{A'}{(\Gamma,\Gamma_i)}{e_i}{A}{e_i'}  \}_i
      \end{tarray} }
    \\
    \ninfer
    {G-BCase}
    { \GenTmp{\Gamma'}{\Delta'}{\BX{\sigma}}{\Gamma}{\CASE~e_0~\OF~\{ p_i \to e_i \}_i}{\sigma}{\BCASE~e_0'~\BOF
        \left\{\begin{tarray}{l}
          \displaystyle \bx{{p_i}} \to e_i'                                         \\
          \displaystyle \BWITH~\EHole{\Gamma'}{\sigma \to \con{Bool}}{e_i'} \\
          \displaystyle \BBY~\RHole{\Gamma'}{\sigma_0 \to \sigma \to \sigma_0}{p_i}{e_i}
        \end{tarray}\right\}_i}}
    { \begin{tarray}{c}
        \GenTmp{\Gamma'}{\Delta'}{\BX{\sigma_0}}{\Gamma}{e_0}{\sigma_0}{e_0'} \\
        \{\Gamma_i\vdash p_i: \sigma_0 \quad
        \Delta'_i \vdash p_i : \sigma_0 \quad
        \GenTmp{\Gamma'}{(\Delta',\Delta_i')}{\BX{\sigma}}{(\Gamma,\Gamma_i)}{e_i}{\sigma}{e_i'}
      \end{tarray} }
  \end{gather*}
  \caption{Sketch generation rules for expressions}
  \label{fig:gen-template}
\end{figure}

\begin{revisedblock}
  \paragraph{{\bf Discussion on \LET-polymorphism}}

  {\SysName} supports $\LET$-polymorphism as illustrated by some $\GET$ programs used in the experiments (Section~\ref{sec:experiment-categorize}), such as $\var{append} :: [a] \to [a] \to [a]$ and $\var{reverse} :: [a] \to [a]$ that have polymorphic types.
  However, for simplicity, we assumed simple types in our formal development (\eg, Fig.~\ref{fig:hobit-typing} and \ref{fig:gen-template}).

  The handling of bidirectional programs with (rank 1) polymorphic types is mostly straightforward. For example, for the sketch generation in Fig.~\ref{fig:gen-template},
  we can leverage the fact that in the standard typing system for $\LET$ polymorphism \cite{DBLP:conf/popl/DamasM82}, expressions can only have monomorphic types.
  As a consequence, we can say $A$ and $A$' in $\GenTmp{\Gamma'}{\Delta'}{A'}{\Gamma}{e}{A}{e'}$ are always monomorphic.
  Also, these types $A$ and $A$' are fully determined in advance of the generation,
  because $A$ comes from the typing derivation of the original $\GET$ program and $A'$ is obtained form $A$ by inserting $\BX{}$. The types $A$ and $A$' may contain
  type variables but they are rigid (\ie, not subject to unification).
  Note that we can access to the typing derivation of the original $\GET$ program as well as the code itself.



  Though it is still straightforward, a slightly more careful discussion is needed for the type-directed generation of terms to fill shape-restricted holes~(Section~\ref{sec:synthesis-filling-generic-holes}), where types being used for type-directed generation may not be immediately clear.
  In an extreme case, for example, if the component functions contains $\var{const} :: a \to b \to a$, we have no immediate information of the type
  to generate its second argument as it cannot be determined by instantiation of $a$.
  This suggests that we may need to consider explicit handing of type variables and unification
  in the term generation, which complicates the system.
  Fortunately though, our back-end system for (lazy) nondeterministic generation is powerful enough to address the issue without introducing any additional complication;
  in the system, a logic (\ie, unifiable) variable (\eg, (non-rigid) type variable) can be expressed by
  nondeterministic generation of all the things (\eg, monomorphic types) to substitute for it,
  and by the $\var{share}$ operator~\cite{DBLP:journals/jfp/FischerKS11}.

\end{revisedblock}

\fi

\ifFullVersion
\subsection{\revised{Full Definition of Filtering based on Branch Traces}}
\label{sec:filtering-traces-full}
\begin{revisedblock}
  Here, we provide details on the filtering based on branch traces, corresponding to Section~\ref{sec:synthesis-prefiltering}.

  A trace is a tree that preserves information about which branches of $\BCASE$ were chosen during the $\GET$ or $\PUT$ evaluation.
  The syntax of such a trace is as follows:
  \[\revised{\mathit{tr}  ::= \epsilon \mid \mathrm{Br}(\mathit{tr}_0,j,\mathit{tr}_1) \mid [\mathit{tr}_1,\ldots,\mathit{tr}_n]}
  \]
  Here, $\mathrm{Br}(\mathit{tr}_0,j,\mathit{tr}_1)$ is a trace for
  $\BCASE~E_0~\BOF~\{ p_i \rightarrow e_i~\WITH~v_i'~\BY~v_i''\}_i$, where $\mathit{tr}_0$ is the trace for $E_0$,
  $j$ stands for the $j$\nolinebreak-th branch that was chosen, and $\mathit{tr}_1$ is the trace for $E_j$.
  The traces $\mathit{tr}_1,\ldots,\mathit{tr}_n$ in
  $[\mathit{tr}_1,\ldots,\mathit{tr}_n]$ correspond to the arguments of a constructor application $\bxcon{C} \A e_1 \A ,\ldots, \A e_n$. We abbreviate $[\mathit{tr}_1,\ldots,\mathit{tr}_n]$ to $\V{\mathit{tr}}$ when $n$ is obvious.

  \paragraph{{\bf Primer on HOBiT's formal semantics}}
  To understand the definition of filtering based on branch traces, one needs to know the formal semantics of HOBiT.
  The semantics of HOBiT is defined so that a term-in-context $\emptyset; \Delta \vdash e : \BX{\sigma}$ defines a bidirectional transformation between $\Delta$ and $\sigma$.
  However, it is not clear how to interpret function abstractions and applications to achieve this goal.
  Thus, HOBiT adopts the staged semantics; it first evaluates $\lambda$s away to obtain a first-order residual expression $E$ from $e$, and
  then interprets $E$ as a bidirectional transformation between $\Delta$ and $\sigma$.
  Specifically, we use three evaluation relations:
  $\EvalRelU{e}{v}$ is for unidirectional evaluation to obtain residual expressions, and
  $\EvalRelF{\mu}{E}{u}$ and $\EvalRelB{\mu}{E}{u}{\mu'}$ for bidirectional evaluation of residual expressions.

  Values $v$, residual expressions $E$ and first-order values $u$ are defined as follows.
  \[
    \begin{tarray}{lcll}
      v &::=& \con{True} \mid \con{False} \mid [\,] \mid v_1 : v_2 \mid \lambda x.e \mid E \\
      E &::=& x \mid  \bx{\con{True}} \mid \bx{\con{False}} \mid \bx{[\,]} \mid E_1 \mathbin{\bx{:}} E_2 \mid \CASE~E_0~\OF~\{ \bx{{p_i}} \to e_i ~\WITH~v'_i~\BY~v''_i \}_i \mid \LIFTU{u} \\
      u &::=& \con{True} \mid \con{False} \mid [\,] \mid u_1 : u_2
    \end{tarray}
  \]
  Intuitively, $v$, $E$, and $u$ represent evaluation results of $A$-, $\BX{\tau}$-, and $\tau$-typed expressions, respectively.
  For more details on HOBiT, please see HOBiT's original paper \cite{Matsuda&Wang:2018:HOBiT}.

  \paragraph{{\bf {\SysName}'s evaluation rules with traces}}
  The rules for the unidirectional evaluation relation are rather standard, as excerpted in Fig.~\ref{fig:hobit-evalU}.
  The bidirectional constructs (\ie, bidirectional constructors and bidirectional $\CASE$) are frozen, \ie, treated as
  ordinary constructors in this evaluation.

  For filtering based on trace branches we make use of the evaluation relations $\EvalRelFT{\mu}{E}{u}{\mathit{tr}}$ and $\EvalRelBT{\mu}{E}{u}{\mathit{tr}}{\mu'}$,
  defined by the rules in Fig.~\ref{fig:get-put-trace-evaluation-rules}.
  Intuitively, $\EvalRelFT{\mu}{E}{v}{\mathit{tr}}$ means that a program $E$ is reduced to the original view $v$ under the the original source environment $\mu$ through a trace $\mathit{tr}$.
  Accordingly, $\EvalRelBT{\mu}{E}{v}{\mathit{tr}}{\mu'}$ means that a program $E$ is evaluated according to its $\PUT$ behavior such that the
  updated source $\mu'$ is obtained from the the original source $\mu$ and the updated view $v$ through a trace $\mathit{tr}$.
  These traced evaluation relations only differ from the original $\GET$ and $\PUT$ evaluation relations~\cite{Matsuda&Wang:2018:HOBiT} in reference to traces.
  Thus, we omit the definitions of the original, untraced evaluation relations $\EvalRelF{\mu}{E}{u}$ and $\EvalRelB{\mu}{E}{u}{\mu'}$, whereas
  they appear in Fig.~\ref{fig:get-put-trace-evaluation-rules}.

  In the evaluation rules, we write $\sem{p}$ for the semantics of pattern matching by $p$: \ie, $\sem{p} \A u$ is a partial function that returns a binding $\mu$
  with $\dom(\mu) = \fv(p)$ and $u = p\mu$ if such $\mu$ exists (and fails otherwise).
  In HOBiT, $\sem{p}$ is injective and thus has a left inverse $\sem{p}^{-1}$.
  The rules for $\BCASE$ (\ie, \rname{Tg-BCase}, \rname{Tp-BCase1}, and \rname{Tp-BCase2}) implicitly assume that $\dom(\mu_i)$ (\ie, $\fv(p_i)$) and $\dom(\mu)$ are disjoint;
  we assume appropriate $\alpha$-renaming that is consistent in $\GET$ and $\PUT$ to fulfill the condition.
  To separate environments, the evaluation rules use $\mu_1 \uplus_{X,Y} \mu_2$ which behaves similarly to $\mu_1 \uplus \mu_2$, while also ensuring $\dom(\mu_1) \subseteq X$ and $\dom(\mu_2) \subseteq Y$.
  The operator $\triangleleft$ is defined by:
  \[
    (\mu' \triangleleft \mu)(x) = \begin{cases}
      \mu'(x)  & \text{if}~x \in \dom(\mu')                      \\
      \mu(x) & \text{if}~x \in \dom(\mu) \setminus \dom(\mu')
    \end{cases}
  \]
  The operator $\lub$ is defined as: $\mu_1 \lub \mu_2 = \mu_1 \cup \mu_2$ if $\mu_1(x) = \mu_2(x)$ for all $x \in \dom(\mu_1) \cap \dom(\mu_2)$,
  and otherwise $\mu_1 \lub \mu_2$ is undefined.

  \begin{figure}[t]
    \setlength{\abovedisplayskip}{0pt}
    \setlength{\belowdisplayskip}{0pt}
    \setlength{\jot}{3pt}
    \begin{revisedblock}
      \begin{gather*}
        \infer
        {\EvalRelU{x}{x}}{}
        \quad
        \infer
        {\EvalRelU{f}{v}}
        {(f = e) \in P \quad \EvalRelU{e}{v}}
        \quad
        \infer
        {\EvalRelU{e_1 \A e_2}{v}}
        {\EvalRelU{e_1}{\lambda x.e} \quad
          \EvalRelU{e_2}{v_2} \quad
          \EvalRelU{e[v_2/x]}{v}
        }
        \quad
        \infer
        {\EvalRelU{\lambda x.e}{\lambda x.e}}
        {}
        \\
        \infer
        {
          \EvalRelU{\CASE~e_0~\OF~\{ \bx{{p_i}} \to e_i ~\WITH~e'_i~\BY~e''_i \}_i}{\CASE~E_0~\OF~\{ \bx{{p_i}} \to e_i ~\WITH~v'_i~\BY~v''_i \}_i}
        }
        {
          \EvalRelU{e_0}{E_0} \quad
          \{ \EvalRelU{e'_i}{v'_i} \quad \EvalRelU{e''_i}{v''_i} \}_i
        }
      \end{gather*}
    \end{revisedblock}
    \caption{\revised{Unidirectional evaluation rules (excerpt)}}
    \label{fig:hobit-evalU}
  \end{figure}

  \begin{figure}
    \setlength{\abovedisplayskip}{0pt}
    \setlength{\belowdisplayskip}{0pt}
    \setlength{\jot}{10pt}
    \footnotesize%
    \begin{revisedblock}
      \begin{gather*}
        \begin{array}{c@{\quad}c@{\quad}c@{\quad}c}
          \displaystyle
          \ninfer{Tg-Var}
          {\EvalRelFT{\mu}{x}{\mu(x)}{\TrNil} }
          {}
           &
          \displaystyle
          \ninfer{Tg-BCon}
          { \EvalRelFT{\mu}{\bxcon{C} \A \V{e}}{\con{C} \A \V{u}}{\V{\mathit{tr}}} }
          { \{\EvalRelFT{\mu}{e_i}{u_i}{\mathit{tr}_i} \}_i}
           &
          \displaystyle
          \ninfer{Tg-!}
          { \EvalRelFT{\mu}{ \LIFTU{u} }{ u }{\TrNil} }
          { }
          \\[10pt]
          \displaystyle
          \ninfer{Tp-Var}
          {\EvalRelBT{\mu}{x}{u}{\TrNil}{ \{x = u\} } }
          {}
           &
          \displaystyle
          \ninfer{Tp-BCon}
          { \EvalRelBT{\mu}{ \bxcon{C} \A \V{e} }{ \con{C} \A \V{u} }{\V{\mathit{tr}}}{\bigcurlyvee \mu'_i } }
          { \{\EvalRelBT{\mu}{e_i}{u_i}{\mathit{tr}_i}{\mu'_i}\}_i }
           &
          \displaystyle
          \ninfer{Tp-!}
          { \EvalRelBT{\mu}{ \LIFTU{u} }{ u }{\TrNil}{\emptyset} }
          {  } \\[10pt]
        \end{array}\\
        \ninfer{Tg-BCase}
        { \EvalRelFT{\mu}{\BCASE~E_0~\OF~\{ \bx{{p_i}} \to e_i ~\BWITH~v'_i~\BBY~v''_i \} }{ u }{ \TrBranch{\mathit{tr}_0}{i}{\mathit{tr}_i} } }
        {
          \EvalRelFT{\mu}{E_0}{u_0}{\mathit{tr}_0} \quad
          \sem{p_i}\A u_0 = \mu_i \quad
          \EvalRelU{e_i}{E_i} \quad
          \EvalRelFT{\mu \uplus \mu_i}{E_i}{u}{\mathit{tr}_i} \quad
          \EvalRelU{v'_i \A u}{\con{True}}
        }\\
        \ninfer{Tp-BCase1}
        { \EvalRelBT{\mu}{\BCASE~E_0~\OF~\{ \bx{{p_i}} \to e_i ~\BWITH~v'_i~\BBY~v''_i \} }{ u }{ \TrBranch{\mathit{tr}_0}{i}{\mathit{tr}_i} }{ \mu'_0 \lub \mu' } }
        {
          \begin{array}{c}
            \EvalRelF{\mu}{E_0}{u_0} \quad
            \sem{p_i}\A u_0 = \mu_i \quad
            \EvalRelU{v'_i \A u}{\con{True}} \\
            \EvalRelU{e_i}{E_i} \quad
            \EvalRelBT{\mu \uplus \mu_i}{E_i}{u}{\mathit{tr}_i}{\mu' \uplus_{\dom(\mu),\dom(\mu_i)} \mu'_i} \quad
            \sem{p_i}^{-1} \A (\mu'_i \triangleleft \mu_i) = u_0' \quad
            \EvalRelBT{\mu}{E_0}{u'_0}{\mathit{tr}_0}{\mu'_0}
          \end{array}
        }\\
        \ninfer{Tp-BCase2}
        { \EvalRelBT{\mu}{\BCASE~E_0~\OF~\{ \bx{{p_i}} \to e_i ~\BWITH~v'_i~\BBY~v''_i \} }{ u }{ \TrBranch{\mathit{tr}_0}{j}{\mathit{tr}_i} }{ \mu'_0 \lub \mu' } }
        {
          \begin{array}{c}
            \EvalRelF{\mu}{E_0}{u_0} \quad
            \sem{p_i}\A u_0 = \mu_i \quad
            \EvalRelU{v'_i \A u}{\con{False}} \quad
            \EvalRelU{v'_j \A u}{\con{True}} \quad
            \{ \EvalRelU{v'_k \A u}{\con{False}} \}_{k < j}
            \\
            \EvalRelU{v''_j \A u_0 \A u}{u_0^\mathrm{rec}} \quad
            \sem{p_j} \A u_0^\mathrm{rec} = \mu_j
            \\
            \EvalRelU{e_j}{E_j} \quad
            \EvalRelBT{\mu \uplus \mu_j}{E_j}{u}{ \mathit{tr}_i }{\mu' \uplus_{\dom(\mu),\dom(\mu_j)} \mu'_j} \quad
            \sem{p_j}^{-1} \A (\mu'_j \triangleleft \mu_j) = u_0' \quad
            \EvalRelBT{\mu}{E_0}{u'_0}{ \mathit{tr}_0 }{\mu'_0}
          \end{array}
        }
      \end{gather*}
    \end{revisedblock}
    \caption{\revised{$\GET$ and $\PUT$ evaluation rules with traces}}
    \label{fig:get-put-trace-evaluation-rules}
  \end{figure}

  For illustration, we use $\var{appendB}$, and assume the following example was given as the backward behavior:
  \[
    \texttt{:put} \A (\var{uncurryB}~ \var{appendB}) \A ([1,2,3], [4,5]) \A [6,2]  = ([6,2], [\,])
  \]
  The trace of
  \[
    \texttt{:get} \A (\var{uncurryB}~ \var{appendB}) \A ([6,2], [\,])  = [6,2]
  \]
  and the trace of the above $\texttt{:put}$ must coincide.
  The trace of this $\texttt{:get}$ (after evaluating $\var{uncurryB}$) is obtained as follows:
  \[ \EvalRelFT{\{\var{xs} = [6,2], \var{ys} = [\,]\}}{\var{appendBbody}}{[6,2]}{\mathrm{Br}(\epsilon,1,[\epsilon,\mathrm{Br}(\epsilon,1,[\epsilon,\mathrm{Br}(\epsilon,0,\epsilon)])])}\]
  where
  \begin{codemath}\bb
    \var{appendBbody} = \BCASE~\var{xs}~\BOF~
    \bbt
    [\,]  & \to \var{ys} & \BWITH~\var{const} \A \con{True} & \BBY~\lambda \dontcare.\lambda \dontcare.\, [\,] \\
    a : x & \to a \bxop{:} \var{appendB} \A x \A \var{ys} & \BWITH~\var{not} \circ \var{null} & \BBY~\lambda s.\lambda \dontcare.\, s 
    \ee
    \ee&&\end{codemath}
  Here, the resulting overall trace is $\mathit{tr} = \mathrm{Br^1}(\epsilon^2,1,[\epsilon^3,\mathrm{Br^4}(\epsilon^5,1,[\epsilon^6,\mathrm{Br^7}(\epsilon^8,0,\epsilon^9)])])$ (for easy identification, we numbered the individual traces). Next, we explain how this is obtained.
  The first step of the evaluation of $\GET$ is the case branch.
  Since we have $\var{xs} = [6,2]$, the second branch (cons case) is chosen.
  Thus, by \rname{Tg-BCase}, the whole trace $\var{tr}$ has the form of $\mathrm{Br^1}(\var{tr}_2, 1, \var{tr}_3)$, 
  where $\var{tr}_2$ is the trace of the evaluation of $\var{xs}$, and $\var{tr}_3$ is the trace of the evaluation of $a \bxop{:} \var{appendB} \A x \A \var{ys}$ under $\{a = 6, x = [2],  \var{ys} = [\,], \dots\}$.
  %
  By \rname{Tg-Var}, we have $\var{tr}_2 = \epsilon^2$, and
  by \rname{Tg-BCon}, we have $\var{tr}_3 = [\var{tr}_4, \var{tr}_5]$ for some trace $\var{tr}_4$ and $\var{tr}_5$.
  Since $\var{tr}_4$ is the trace of the evaluation of $\var{a}$, we have $\var{tr}_4 = \epsilon^3$ by \rname{Tg-Var}.
  Then, we focus on $\var{tr}_5$, which is the trace of the evaluation of $\var{appendB} \A x \A \var{ys}$ under $\{x = [2], \var{ys} = [\,], \dots\}$.
  Thus, we have\footnote{\label{footnote:abused}If we follow the evaluation rules precisely, we will evaluate $\var{appendBbody}[x/\var{xs}, \var{ys}/\var{ys}]$ with $\alpha$-renaming of $a$ and $x$ bound in $\var{appendBbody}$, which would bring further complication and thus is ignored here.}
  \[
    \EvalRelFT{\{\var{xs} = [2], \var{ys} = [\,], \dots\}}{\var{appendBbody}}{v_5}{\var{tr}_5}
  \]
  for some value $v_5$.
  The second branch (cons case) is again chosen, and we obtain $\var{tr}_5 = \mathrm{Br^4}(\epsilon^5,1,[\epsilon^6,\mathit{tr}_6])$ for some trace $\mathit{tr}_6$.
  Since $\var{tr}_6$ is the trace of the evaluation of $\var{appendB} \A x \A \var{ys}$ under $\{x = [\,], \var{ys} = [\,], \dots\}$, we have
  \[
    \EvalRelFT{\{\var{xs} = [\,], \var{ys} = [\,], \dots\}}{\var{appendBbody}}{v_6}{\var{tr}_6}
  \]
  for some value $v_6$.
  Since we have $\var{xs} = [\,]$, the first branch (nil case) is chosen. Thus, by \rname{Tg-BCon}, we have $\var{tr}_6 = \mathrm{Br^7}(\var{tr}_7, 0, \var{tr}_8)$ for some $\var{tr}_7$ and $\var{tr}_8$.
  Since $\var{tr}_7$ is the trace of the evaluation of $\var{xs}$, we have $\var{tr}_7 = \epsilon^8$ by \rname{Tg-Var}.
  By \rname{Tg-Var}, we have $\var{tr}_8 = \epsilon^9$.
  As a result, the whole trace of the $\GET$ direction is computed as:
  \[\mathit{tr} = \mathrm{Br^1}(\epsilon^2,1,[\epsilon^3,\mathrm{Br^4}(\epsilon^5,1,[\epsilon^6,\mathrm{Br^7}(\epsilon^8,0,\epsilon^9)])])\]

  Then, we check if $\texttt{:put} \A (\var{uncurryB}~ \var{appendB}) \A ([1,2,3], [4,5]) \A [6,2] $ goes through the same trace $\mathit{tr}$ as $\GET$.
  Essentially, this means that the following relation must hold:
  \[
    \EvalRelBT{\{\var{xs} = [1,2,3], \var{ys} = [4,5]\}}{\var{appendBbody}}{[6,2]}{\mathrm{Br}(\epsilon,1,[\epsilon,\mathrm{Br}(\epsilon,1,[\epsilon,\mathrm{Br}(\epsilon,0,\epsilon)])])}{\mu'}
    \tag{\textdagger}\label{prop:goal}
  \]
  with $\mu' = \{ \var{xs} = [6,2], \var{ys} = [\,] \}$.
  In this evaluation, we also make use of the trace to determine the rule to be applied.
  For example, in the backward evaluation of $\var{appendBbody}$ above, we know from the trace that the second branch (cons case) is chosen.
  Also, we know from the original environment, which maps $\var{xs}$ to $[1,2,3]$, that branch switching does not happen.
  Thus, the last rule used in the evaluation must be \rname{Tp-BCase1}.
  Then, we need to check the following conditions that appear as the premises of the rule (we shall omit the $\GET$ evaluation of scrutinee and the obvious pattern matching as they are not relevant here).
  \begin{gather*}
    \EvalRelU{(\var{not} \circ \var{null}) \A [6,2]}{\con{True}}\\
    \EvalRelBT{ \{a = 1, x = [2,3], \var{ys} = [4,5], \dots\} }{ a \bxop{:} \var{appendB} \A x \A \var{ys} }{ [6,2] }{\var{tr}_3}{ \mu'_1 \uplus_{\{\var{ys}, \dots\}, \{a,x\}} \mu'_2 } \tag{\ensuremath{\ast}}\label{prop:put1}\\
    \EvalRelBT{ \{ \var{xs} = [1,2,3], \dots \} }{ \var{xs} }{ \sem{a : x}^{-1} (\mu_2' \triangleleft \{ a = 1, x = [2,3] \}) }{\var{tr}_2}{ \mu'_3 }\\
    \mu' = \mu'_1 \lub \mu'_3
  \end{gather*}
  Here, $\var{tr}_2, \var{tr}_3, \dots$ are the traces that appeared in the $\GET$ evaluation; hence $\var{tr}_2 = \epsilon$ and $\var{tr}_3 = [\var{tr}_4, \var{tr}_5] = [\epsilon,\mathrm{Br}(\epsilon,1,[\epsilon,\mathrm{Br}(\epsilon,0, \epsilon)])]$ for example.
  The first condition clearly holds, and the third condition can be solved by \rname{Tp-Var} with
  $\mu'_3 = \{ \var{xs} = \sem{a : x}^{-1} (\mu_2' \triangleleft \{ a = 1, x = [2,3] \}) \}$. So, let's focus on the
  second condition (\ref{prop:put1}).
  By \rname{Tp-BCon}, this condition is reduced to
  \begin{gather*}
    \EvalRelBT{ \{a = 1, \dots\} }{ a  }{ 6 }{\var{tr}_4}{ \mu'_4 }
  \end{gather*}
  and checking whether $\var{tr}_5$ is the trace of the backward evaluation of $\var{append} \A x \A \var{ys}$
  under $\{ x = [2,3], \var{ys} = [4,5], \dots \}$ for the updated view $[2]$ to yield $\mu'_5$ such that $\mu'_1 \uplus_{\{\var{xs},\var{ys}\}, \{a,x\}} \mu'_2  = \mu'_4 \lub \mu'_5$.
  By \rname{Tp-Var}, it is easy to see that the former holds with $\mu'_4 = \{ a = 6 \}$.
  The latter holds if (and only if)\footnote{Similar to Footnote~\ref{footnote:abused}.}
  \[
    \EvalRelBT{\{\var{xs} = [2,3], \var{ys} = [4,5], \dots \}}{\var{appendBbody}}{[2]}{\var{tr}_5}{\mu'_6}
  \]
  holds with $\var{tr}_5 = \mathrm{Br}(\epsilon,1,[\epsilon, \var{tr}_6]) = \mathrm{Br}(\epsilon, 1,[\epsilon,\mathrm{Br}(\epsilon,0,\epsilon)])$ and $\mu'_5 = \{ x = \mu'_6(\var{xs}), \var{ys} = \mu'_6(\var{ys})\}$.
  Again, in the backward evaluation of $\var{appendBbody}$, we know from the trace $\var{tr}_5$ that the second branch (cons case) must be chosen, and from the original environment $\{ \var{xs} = [2,3], \dots \}$ that the branch is the original one. Thus, the last rule used for the evaluation must be \rname{Tp-BCase1}.
  Thus, to make the above traced backward evaluation hold, it suffices (and needs) that the following conditions hold (again, we shall ignore irrelevant premises of \rname{Tp-BCase1}).
  \begin{gather*}
    \EvalRelU{(\var{not} \circ \var{null}) \A [2] }{ \con{True} } \\
    \EvalRelBT{ \{a = 2, x = [3], \var{ys} = [4,5], \dots\} }{ a \bxop{:} \var{appendB} \A x \A \var{ys} }{ [2] }{[\epsilon, \var{tr}_6]}{ \mu'_7 \uplus_{\{\var{ys}, \dots\}, \{a,x\}} \mu'_8 } \tag{\ensuremath{\ast\ast}}\label{prop:put2}\\
    \EvalRelBT{ \{ \var{xs} = [2,3], \dots \} }{ \var{xs} }{ \sem{a : x}^{-1} (\mu_8' \triangleleft \{ a = 2, x = [3] \}) }{\epsilon}{ \mu'_9 }  \\
    \mu'_6 = \mu'_7 \lub \mu'_9
  \end{gather*}
  The first condition clearly holds, and the third condition can be solved by \rname{Tp-Var} with
  $\mu'_9 = \{ \var{xs} = \sem{a : x}^{-1} (\mu_8' \triangleleft \{ a = 2, x = [3] \})$.
  The second condition (\ref{prop:put2}) is further reduced to conditions:
  \begin{gather*}
    \EvalRelBT{ \{a = 2, \dots\} }{ a  }{ 2 }{\epsilon}{ \mu'_{10} } \\
    \EvalRelBT{\{\var{xs} = [3], \var{ys} = [4,5], \dots \}}{\var{appendBbody}}{[\,]}{\var{tr}_6}{\mu'_{11}}\\
     \mu'_7 \uplus_{\{\var{ys}, \dots\}, \{a,x\}} \mu'_8 = \mu'_{10} \lub \{ \var{x} = \mu'_{11}(\var{xs}), \var{ys} = \mu'_{11}(\var{ys}) \}\text{.}
  \end{gather*}
  Clearly, by \rname{Tp-Var}, the first condition holds with $\mu'_{10} = \{ a = 2 \}$.
  Thus, let's focus on the second condition.
  Unlike the previous cases, the trace $\var{tr}_6 = \mathrm{Br}(\epsilon,0,\epsilon)$ says that the evaluation must use the first branch (nil case),
  which is different from the original branch (taken under the environment $\{ \var{xs} = [3], \dots \}$).
  Thus, the last rule used in the evaluation
  \(
  \EvalRelBT{\{\var{xs} = [3], \var{ys} = [4,5], \dots \}}{\var{appendBbody}}{[\,]}{\var{tr}_6}{\mu'_{11}}
  \)
  must be \rname{Tp-BCase2}, meaning that branch switching happened. Since we have
  \[
    \EvalRelU{ (\lambda \dontcare. \lambda \dontcare.\, [\,]) \A [3] \A [\,] }{ [\,] }
  \]
  we then check other premises of the rule instance (again, we shall ignore the irrelevant ones):
  \begin{gather*}
    \EvalRelU{(\var{not} \circ \var{null}) \A [\,]}{\con{False}} \\
    \EvalRelU{(\var{const} \A \con{True}) \A [\,]}{\con{True}} \\
    \EvalRelBT{ \{ \var{ys} = [4,5], \dots \} }{ \var{ys} }{ [\,] }{ \epsilon }{ \mu'_{12} \uplus_{\{ \var{ys}, \dots\}, \emptyset} \emptyset} \\
    \EvalRelBT{ \{ \var{xs} = [3], \dots \}  }{ \var{xs} }{ \sem{ [\,] }^{-1}(\emptyset \triangleleft \emptyset) }{ \epsilon }{ \mu'_{13} }\\
    \mu'_{11} = \mu'_{12} \lub \mu'_{13}\text{.}
  \end{gather*}
  The first two clearly hold, and by \rname{Tp-Var}, the third and forth ones hold with  $\mu'_{12} = \{ \var{ys} = [\,] \}$ and $\mu'_{13} = \{ \var{xs} = [\,] \}$.
  Thus, we have $\mu'_{11} = \{ \var{xs} = [\,], \var{ys} = [\,] \}$ to satisfy the fifth condition.
  Then, we can solve the constraints on environments obtained so far as:
  $\mu'_8 = \{ a = 2, x = [\,] \}$,
  $\mu'_7 = \{ \var{ys} = [\,] \}$,
  $\mu'_9 = \{ \var{xs} = [2] \}$,
  $\mu'_6 = \{ \var{xs} = [2], \var{ys} = [\,] \}$,
  $\mu'_5 = \{ x = [2], \var{ys} = [\,] \}$,
  $\mu'_1 = \{ \var{ys} = [\,] \}$,
  $\mu'_2 = \{ a = 6, x = [2] \}$, and
  $\mu'_3 = \{ \var{xs} = [6,2] \}$. Now, we are ready to check $\mu' = \{ \var{xs} = [6,2], \var{ys} = [\,] \} = \mu'_1 \lub \mu'_3$ to conclude (\ref{prop:goal}).

\end{revisedblock}
\fi

\ifFullVersion

\subsection{\revised{A Concrete Input and Output for Q1}}
\label{sec:concrete-input-output-for-q1}
\begin{revisedblock}




To demonstrate that \SysName{} is able to generate relatively large and complex (for automatic program synthesis) programs,
we give here the input specification and synthesized output corresponding to Q1 in our experiments (Table~\ref{fig:result-experiments-xml} in Section~\ref{sec:XML_Transfromation}).
Figures~\ref{fig:XML_example_get}, \ref{fig:XML_example_original_source}, \ref{fig:XML_example_updated_view} and \ref{fig:XML_example_updated_source} represent the input specification given to {\SysName} and Figure~\ref{fig:XML_example_output} is
the corresponding output. Here, we used the rose tree datatype $\con{Tree}$ (Fig.~\ref{fig:XML_example_get}) to express XML fragments.
Note that the output involves non-trivial exit conditions and reconciliation functions.



  \begin{figure}[t]
    \begin{revisedblock}
\begin{Verbatim}
data Tree a = N a [Tree a]  -- polymorphic tree type

data Lab = A [Char] [Char]  -- Attribute
         | T [Char]         -- Text
         | E [Char]         -- Element

q1 :: Tree Lab -> Tree Lab
q1 t = let N (E "book") ts = t
       in N (E "toc") (q1_section ts)

q1_section :: [Tree Lab] -> [Tree Lab]
q1_section l = case l of
  []                         -> []
  N (A a b) [] : rest        -> N (A a b) [] : q1_section rest
  N (E "title") title : rest -> N (E "title") title : q1_section rest
  N (E "section") xs : rest  -> N (E "section") (q1_section xs) : q1_section rest
  node:rest                  -> q1_section rest
\end{Verbatim}
    \end{revisedblock}
    \caption{\revised{Input $\GET$ program}}
    \label{fig:XML_example_get}
  \end{figure}

  \begin{figure}[t]
    \begin{revisedblock}
\begin{Verbatim}
N
  (E "book")
  [ N (E "title") [N (T "Data on the Web") []]
  , N (E "author") [N (T "Serge Abiteboul") []]
  , N (E "author") [N (T "Peter Buneman") []]
  , N (E "author") [N (T "Dan Suciu") []]
  , N
      (E "section")
      [ N (A "id" "intro") []
      , N (A "difficulty" "easy") []
      , N (E "title") [N (T "Introduction") []]
      , N (E "p") [N (T "Text ... ") []]
      , N
          (E "section")
          [ N (E "title") [N (T "Audience") []]
          , N (E "p") [N (T "Text ... ") []]]
      , N
          (E "section")
          [ N (E "title") [N (T "Web Data and the Two Cultures") []]
          , N (E "p") [N (T "Text ... ") []]
          , N
              (E "figure")
              [ N (A "height" "400") []
              , N (A "width" "400") []
              , N
                  (E "title")
                  [N (T "Traditional client/server architecture") []]
              , N (E "image") [N (A "source" "csarch.gif") []]]
          , N (E "p") [N (T "Text ...") []]]]
  , N
      (E "section")
      [ N (A "id" "syntax") []
      , N (A "difficulty" "medium") []
      , N (E "title") [N (T "A Syntax For Data") []]
      , N (E "p") [N (T "Text ... ") []]
      , N
          (E "figure")
          [ N (A "height" "200") []
          , N (A "width" "500") []
          , N (E "title") [N (T "Graph representations of structures") []]
          , N (E "image") [N (A "source" "graphs.gif") []]]
      , N (E "p") [N (T "Text ... ") []]
      , N
          (E "section")
          [ N (E "title") [N (T "Base Types") []]
          , N (E "p") [N (T "Text ...") []]]
      , N
          (E "section")
          [ N (E "title") [N (T "Representing Relational Databases") []]
          , N (E "p") [N (T "Text") []]
          , N
              (E "figure")
              [ N (A "height" "250") []
              , N (A "width" "400") []
              , N (E "title") [N (T "Examples of Relations") []]
              , N (E "image") [N (A "source" "relatios.gif") []]]]
      , N
          (E "section")
          [ N (E "title") [N (T "Representing Object Databases") []]
          , N (E "p") [N (T "Text ... ") []]]]]
\end{Verbatim}
    \end{revisedblock}
    \caption{\revised{Original source}}
    \label{fig:XML_example_original_source}
  \end{figure}

  \begin{figure}[t]
    \begin{revisedblock}
\begin{Verbatim}
N
  (E "toc")
  [ N (E "title") [N (T "Data on the Web") []]
  , N
      (E "section")
      [ N (A "id" "intro") []
      , N (A "newattr" "attr") []  -- added
      , N (A "difficulty" "easy") []
      , N (E "title") [N (T "Introduction") []]
      , N (E "section") [N (E "title") [N (T "Audience") []]]]
      --   , N
      --       (E "section")
      --       [N (E "title") [N (T "Web Data and the Two Cultures") []]]
  , N
      (E "section")
      [ N (A "id" "syntax") []
      , N (A "difficulty" "hard") [] -- easy -> hard
      , N (E "title") [N (T "A Syntax For Data") []]
      , N (E "section") [N (E "title") [N (T "Base Types") []]]
      , N
          (E "section")
          [ N
              (E "title")
              [N (T "Representing Relational Databases and so on") []]]  -- changed title
      , N
          (E "section")
          [N (E "title") [N (T "Representing Object Databases") []]]
      , N (E "section") [N (E "title") [N (T "new section") []]]]  -- added
  , N (E "section") [N (E "title") [N (T "new section") []]]]      -- added
\end{Verbatim}
    \end{revisedblock}
    \caption{\revised{Updated view (with comments added to denote the changes from an original view)}}
    \label{fig:XML_example_updated_view}
  \end{figure}

  \begin{figure}[t]
    \begin{revisedblock}
\begin{Verbatim}
N
  (E "book")
  [ N (E "title") [N (T "Data on the Web") []]
  , N (E "author") [N (T "Serge Abiteboul") []]
  , N (E "author") [N (T "Peter Buneman") []]
  , N (E "author") [N (T "Dan Suciu") []]
  , N
      (E "section")
      [ N (A "id" "intro") []
      , N (A "newattr" "attr") []  -- added
      , N (A "difficulty" "easy") []
      , N (E "title") [N (T "Introduction") []]
      , N (E "p") [N (T "Text ... ") []]
      , N
          (E "section")
          [ N (E "title") [N (T "Audience") []]
          , N (E "p") [N (T "Text ... ") []]]]
          -- , N
          --     (E "section")
          --     [ N (E "title") [N (T "Web Data and the Two Cultures") []]
          --     , N (E "p") [N (T "Text ... ") []]
          --     , N
          --         (E "figure")
          --         [ N (A "height" "400") []
          --         , N (A "width" "400") []
          --         , N
          --             (E "title")
          --             [N (T "Traditional client/server architecture") []]
          --         , N (E "image") [N (A "source" "csarch.gif") []]]
          --     , N (E "p") [N (T "Text ...") []]]]
  , N
      (E "section")
      [ N (A "id" "syntax") []
      , N (A "difficulty" "hard") []  -- easy -> hard
      , N (E "title") [N (T "A Syntax For Data") []]
      , N (E "p") [N (T "Text ... ") []]
      , N
          (E "figure")
          [ N (A "height" "200") []
          , N (A "width" "500") []
          , N (E "title") [N (T "Graph representations of structures") []]
          , N (E "image") [N (A "source" "graphs.gif") []]]
      , N (E "p") [N (T "Text ... ") []]
      , N
          (E "section")
          [ N (E "title") [N (T "Base Types") []]
          , N (E "p") [N (T "Text ...") []]]
      , N
          (E "section")
          [ N
              (E "title")
              [N (T "Representing Relational Databases and so on") []]  -- changed title
          , N (E "p") [N (T "Text") []]
          , N
              (E "figure")
              [ N (A "height" "250") []
              , N (A "width" "400") []
              , N (E "title") [N (T "Examples of Relations") []]
              , N (E "image") [N (A "source" "relatios.gif") []]]]
      , N
          (E "section")
          [ N (E "title") [N (T "Representing Object Databases") []]
          , N (E "p") [N (T "Text ... ") []]]
      , N (E "section") [N (E "title") [N (T "new section") []]]]  -- added
  , N (E "section") [N (E "title") [N (T "new section") []]]]      -- added
\end{Verbatim}
    \end{revisedblock}
    \caption{\revised{Updated source (with comments added to denote the changes from the original source)}}
    \label{fig:XML_example_updated_source}
  \end{figure}

  \begin{figure}[t]
    \begin{revisedblock}
\begin{Verbatim}
q1 :: BX (Tree Lab) -> BX (Tree Lab)
q1 = \t -> case* t of
              N (E "book") ts ->
                  N (E (| "toc" |)) (q1_section ts)
                  with (\x1 -> True)
                  reconciled by (\x0 -> \x1 -> x0)

q1_section :: BX [Tree Lab] -> BX [Tree Lab]
q1_section = \l -> case* l of
                      [] ->
                          ![]
                          with (\x1 -> case x1 of
                                            [] -> True
                                            _ -> False)
                          reconciled by (\x0 -> \x1 -> [])
                      N (A a b) [] : rest ->
                          (|N (A a b) ![] : q1_section rest|)
                          with (\x1 -> case x1 of
                                            N (A x2 x3) [] : x4 -> True
                                            _ -> False)
                          reconciled by (\x0 -> \x1 -> case x1 of
                                                            N (A x2 x3) [] : x4 ->
                                                                N (A x2 x2) [] : x0)
                      N (E "title") title : rest ->
                          (|N (E (| "title" |)) title : q1_section rest|)
                          with (\x1 -> case x1 of
                                            N (E "title") x2 : x3 -> True
                                            _ -> False)
                          reconciled by (\x0 -> \x1 -> case x1 of
                                                            N (E "title") x2 : x3 -> x0)
                      N (E "section") xs : rest ->
                          (|N (E (| [ "section"  ] |)) (q1_section xs) : q1_section rest|)
                          with (\x1 -> case x1 of
                                            N (E "section") x2 : x3 -> True
                                            _ -> False)
                          reconciled by (\x0 -> \x1 -> case x1 of
                                                            N (E "section") x2 : x3 ->
                                                                N (E "section") x2 : x0)
                      node : rest ->
                          q1_section rest
                          with (\x1 -> True)
                          reconciled by (\x0 -> \x1 -> x0)
\end{Verbatim}
    \end{revisedblock}
    \caption{\revised{Output of {\SysName}}}
    \label{fig:XML_example_output}
  \end{figure}
\end{revisedblock}

\clearpage

\fi

\fi

\end{document}